\documentclass[lettersize,journal]{IEEEtran}

\def\revisionthreeclean{}

\usepackage{graphicx}
\usepackage{cite}
\usepackage{CJK}   
\usepackage{amsmath,amssymb}
\usepackage{booktabs}
\usepackage{xurl}
\usepackage{tabu}
\usepackage{tabularx}
\usepackage{multirow}
\usepackage{makecell} 
\usepackage{xcolor}
\usepackage{enumitem}
\usepackage{listings}
\usepackage{soul}
\usepackage{calc}
\usepackage{tikz}
\usepackage{tcolorbox}
\usepackage[hidelinks]{hyperref}

\definecolor{varbg}{HTML}{E8EAED} 
\definecolor{promptbg}{HTML}{F8F9FA}
\definecolor{promptframe}{HTML}{D1D5DB}
\definecolor{CaptionColor}{HTML}{515151}

\newcommand{\revision}[1]{#1}

\definecolor{revisionthreepurple}{HTML}{7545A0}
\ifdefined\revisionthreeclean
  \newcommand{\revisionThree}[1]{#1}
\else
  \newcommand{\revisionThree}[1]{\leavevmode{\textcolor{revisionthreepurple}{#1}}}
\fi

\newcommand{\revisionblock}[1]{#1}
\newcommand{\remark}[1]{}
\newcommand{\removed}[1]{}

\newcommand{\highlight}[1]{\textbf{\underline{#1}}}
\newcommand{\toolname}{\textit{DataRecycle}}

\makeatletter
\let\revisionprevious\revision
\DeclareRobustCommand{\revision}{\@ifnextchar3{\revisionthree@dispatch}{\revisionprevious}}
\def\revisionthree@dispatch3#1{\revisionThree{#1}}
\makeatother

\newcommand*\captionID[1]{\tikz[baseline=(char.base)]{
    \node[
        shape=rectangle,
        fill=black!12,
        text=black,
        inner sep=1.2pt,
        minimum size=8pt,
        rounded corners=1pt
    ] (char) {#1};
}}

\newcommand*\captionNum[1]{\tikz[baseline=(char.base)]{
    \node[
        shape=rectangle,
        fill=white,
        draw=black,
        text=black,
        inner sep=1.2pt,
        minimum size=8pt,
        rounded corners=1pt,
        line width=1pt
    ] (char) {#1};
}}

\lstdefinestyle{beautiful_listing}{
    basicstyle=\ttfamily\small,
    backgroundcolor=\color{promptbg},
    rulecolor=\color{promptframe},
    frame=single,
    framesep=8pt,
    framerule=0.5pt,
    breaklines=true,
    breakatwhitespace=true,
    columns=fullflexible,
    keepspaces=true,
    showstringspaces=false,
    numbers=none,
    escapeinside={(*@}{@*)},
    linewidth=\dimexpr\linewidth-16pt-1pt\relax,
    xleftmargin=0pt,
    xrightmargin=0pt,
    framexleftmargin=0pt,
    framexrightmargin=0pt
}

\setlist[itemize]{
    leftmargin=*,
    itemsep=0pt,
    topsep=2pt,
    parsep=0pt,
    partopsep=0pt
}

\title{DataRecycle: Trustworthy Distillation of Structured Data from Scattered Web Evidence}

\author{%
  Can Liu,
  Tengteng Zhu,
  Xinyu Jiang,
  Xiaolin Wen,
  Yong Wang
  \thanks{C. Liu is with the School of Information and the Institute for the Community for the Chinese Nation at the Central University of Finance and Economics, and with Nanyang Technological University.
  This work was conducted primarily while he was a research fellow supervised by Yong Wang at Nanyang Technological University.
  Email: can.liu@cufe.edu.cn.}
  \thanks{X. Wen and Y. Wang are with Nanyang Technological University. Email: xiaolin004@e.ntu.edu.sg, yong-wang@ntu.edu.sg.}
  \thanks{Y. Wang is the corresponding author.}
  \thanks{T. Zhu is with Beijing Institute of Technology and Nanyang Technological University. Email: bit\_zhutengteng@163.com.}
  \thanks{X. Jiang is with the University of Sydney. Email: xjia0751@uni.sydney.edu.au}
}
\begin{document}
\begin{CJK}{UTF8}{gbsn}  

\maketitle

\begin{abstract}
\revision{The web contains vast amounts of scattered quantitative evidence that is difficult to organize into comparable and verifiable structures.}
Yet its unstructured, sparse, and sometimes conflicting nature makes a systematic distillation costly and error-prone.
Existing approaches force users to make a trade-off: search engines return document lists that require labor-intensive extraction and reconciliation, while generative AI systems produce fluent summaries that obscure data provenance, introduce reporting bias, and hide inconsistencies.
\revision{We present \toolname{}, an interactive system that helps transform scattered web evidence into structured tables of comparable numerical facts.}
Given a user's query or statement, \toolname{} decomposes it into entities and attributes, retrieves candidate evidence, and organizes scattered, potentially conflicting records in a table-based evidence view.
To address key challenges in automated data distillation (i.e., incomplete retrieval, untraceable generation, sparse data distribution, and multi-source inconsistency),
\toolname{} adopts an ontology-guided, top-down data acquisition strategy, visualizes the recorded lineage from retrieved snippets to extracted values, and aligns heterogeneous data across dimensions such as time, definition, and scale.
\revision{We evaluate \toolname{} through module-level experiments, a usage scenario, and a user study, showing that the system helps users organize numerical web evidence into provenance-linked tables and inspect uncertainty and conflicting sources.}

\end{abstract}

\begin{IEEEkeywords}
Data aggregation, uncertainty visualization, fact verification
\end{IEEEkeywords}


\section{Introduction}



While existing curated datasets provide a solid foundation for systematic analysis, they often cover only a limited portion of the knowledge available in the world~\cite{machine_knowledge}. Much of the most timely and fine-grained data, such as news, corporate disclosures, and market reports, first appears as unstructured text on web pages~\cite{web_data_extraction} before being incorporated into official datasets.
Converting such evidence into official structured datasets requires substantial time and effort, leading to an inevitable lag between real-world events and dataset updates. Consequently, to capture emerging trends, underrepresented sectors, or the long tail of entities and metrics outside the scope of major data providers, researchers often need to turn to the web as a data source.

Despite this potential, much of the available web data remains underutilized because it is largely unstructured.
Unlike structured datasets, in which entities, attributes, and relationships are explicitly defined, valuable quantitative facts on the web are scattered across different websites.
These facts are typically buried within unstructured text, which makes systematic analysis difficult.
For instance, assessing the global green energy landscape requires aggregating data points from diverse sector-specific reports.
\revision{Many such data points are absent from global statistical datasets; even when included, they are often poorly aligned, reported in inconsistent units, or associated with conflicting values because of differences in methodology or reporting cycles.}
\revision{Therefore, distilling comparable structured facts directly from the web is essential for bridging the gap between curated datasets and evidence reported in the wild.}

Current tools largely follow two document-centered workflows: search engines return ranked pages, while generative engines~\cite{aggarwal2024geo} summarize retrieved pages into prose.
\revision{Traditional search engines return ranked lists of web pages, leaving users to navigate multiple pages, locate relevant facts, align them across entities, and manually resolve contradictions.}
A recent system, Compendia~\cite{karunathilaka2026compendia}, partially reduces this burden by generating visual narratives from retrieved documents, but it remains document-centric.
In particular, it does not construct a table of the underlying facts, making fact-level coverage difficult to inspect.
\revision{Generative engines such as Perplexity\footnote{\url{https://www.perplexity.ai/}} and ChatGPT\footnote{\url{https://chatgpt.com/}} improve usability by synthesizing direct answers, but their answers are still shaped by both retrieval coverage and generation-time selection.
Even when relevant pages are retrieved, numerical facts may be selectively summarized, merged into prose, or omitted, making it difficult for users to inspect which entity--attribute cells are supported, missing, or conflicting across sources.}
Beyond verifying individual data points, users often need context to interpret what a numerical value actually means.
This requires the same attributes to be aligned across entities, yet document retrieval often makes such alignment inconsistent because different retrieved pages emphasize different aspects of each entity.
For example, if the top retrieved documents for Country A emphasize installed capacity whereas those for Country B focus on subsidies, a generative engine may produce a narrative that fails to compare the same metric across countries.
Moreover, generative engines often smooth over such inconsistencies and obscure data sparsity.
By collapsing conflicting evidence into a single narrative, they hide the provenance and distribution of the underlying data, making it difficult for users to assess uncertainty, identify contradictions, or determine where data is fundamentally missing.

To address this problem, we first identify the fundamental challenges that hinder the transformation of scattered web fragments into verifiable and structured data (C1-C4).

\begin{itemize}
    \item \textbf{C1. Incompleteness of retrieval results.}
    Bottom-up retrieval and summarization methods based on relevance often fail to cover the full scope of a target schema. Search engines and generative engines cannot guarantee retrieval of all data facts required to construct a complete comparison table.

    \item \textbf{C2. Untraceability of generation process.}
    The data generation process in existing approaches often lacks traceable lineage.
    Users cannot verify how specific values are extracted or computed from source documents, making it difficult to distinguish grounded facts from model hallucinations.

    \item \textbf{C3. Sparsity of data distribution.}
    Quantitative data is often sparsely distributed and isolated across unrelated documents. A critical data point may appear only once in a lengthy report, which makes it difficult to aggregate dispersed fragments into a coherent structure.
    For example, when users compare attributes across multiple countries, relevant evidence may not appear in the same document unless prior analyses have already consolidated it.


    \item \textbf{C4. Inconsistency in multi-source evidence.}
    Data retrieved from heterogeneous sources often contains conflicts.
    These discrepancies arise from both factual errors and mismatches in dimensions, such as differences in time frames, scopes, or attribute definitions, which complicate validation.

\end{itemize}

\noindent \revision{To address these challenges, we propose \toolname{}, an interactive system for trustworthiness-oriented distillation of structured tables of comparable numerical facts from scattered web evidence.}
Given a user query or statement, \toolname{} decomposes it into constituent entities, attributes, and scope, retrieves relevant web sources, and integrates fragmented evidence into a unified view.
To guide retrieval toward schema-level coverage (C1), \toolname{} adopts an ontology-guided top-down approach.
Instead of relying only on relevance-ranked retrieval, which can leave some schema cells under-covered, the system first specifies a target schema and then retrieves and extracts evidence for each entity-attribute cell.
To support traceability (C2), \toolname{} records and visualizes the system-observed lineage from retrieved snippets to final data values, allowing users to inspect the extraction and computation steps represented in the pipeline.
To address data sparsity (C3) and potential conflicts (C4), we introduce a context-aware alignment mechanism that normalizes heterogeneous data points into a unified coordinate system, for example, by standardizing temporal, unit, and attribute dimensions, together with a multidimensional source-quality signal that helps prioritize potentially conflicting evidence for further inspection.
\revision{To further support the inspection of conflicting multi-source evidence (C4), we design a compact uncertainty encoding based on the complementary cumulative distribution function (CCDF) for dense tabular comparisons, summarizing source-supported value distributions while preserving comparability across entities.}
With \toolname{}, users can transform natural-language queries into structured comparison tables, trace evidence for each cell, and inspect conflicting perspectives interactively.

Our main contributions are as follows:

\begin{itemize}
    \item \revision{We present a trustworthiness-oriented data engine for distilling comparable structured facts from scattered web sources, integrating retrieval, evidence extraction, context-aware alignment, and source-aware aggregation.}
    \item \revision{We develop \toolname{}, an interactive system that supports provenance-aware table construction and uncertainty inspection, including a compact CCDF-based encoding for dense comparison views.}
    \item We extensively evaluated \toolname{}: a usage scenario to showcase its workflow, quantitative benchmarking of the evidence extraction and record-level auditing modules, a controlled comparison of CCDF bars and density plots, and a user study on the provenance interface, which collectively demonstrated its effectiveness and usability.
    



\end{itemize}

\section{Related Work}\label{sec:related}
This work is related to prior research on web data extraction, data trustworthiness, and uncertainty visualization.

\subsection{Web Data Extraction}

Prior work on web data extraction mainly follows two paradigms: automated extraction and structure discovery with constraint modeling.

\textbf{Automated Extraction.}
Early work focused on scalable information extraction using predefined patterns or relation-learning models, such as Snowball~\cite{agichtein2000snowball} and OpenIE~\cite{banko2007open}. 
Subsequent systems shifted toward probabilistic and declarative frameworks, including DeepDive~\cite{zhang2017deepdive} and Knowledge Vault~\cite{dong2014knowledge}, which integrate heterogeneous signals for large-scale knowledge base construction. 
These approaches support high-throughput extraction, but they typically assume relatively stable schemas or rely on post hoc normalization.
ChartifyText~\cite{zhang2024chartifytext} uses an LLM to infer tables from text containing numerical data and generate charts from those tables.

\textbf{Structure Discovery and Constraint Modeling.}
Beyond direct extraction, recent studies have investigated how to uncover latent structures from noisy, heterogeneous text.
Structure profiling techniques use clustering and representation learning, for example, StructVizor~\cite{huang2025structvizor} and Tsai et al.~\cite{tsai2025harmful}, or graph-based formulations such as clique-expansion hypergraphs in HINTs~\cite{lee2025hints}, to identify implicit schemas.
In parallel, constraint-guided methods improve consistency by incorporating domain knowledge, including state-based decoding~\cite{li2025bridging}, in-context log encoding~\cite{ni2025fgcibgc}, purpose-driven inference~\cite{chen2025semantics}, and hierarchical blueprint representations~\cite{chen2025conch, ferreira2025vablueprint}. 
These methods improve structural coherence, but they remain largely automated and opaque to end users.
\revision3{SciDaSynth~\cite{wang2025scidasynth} lets users query a screened set of scientific papers, check extracted values against the papers, and refine the table.}

\revision3{Existing systems such as ChartifyText and SciDaSynth structure information from supplied texts or screened collections of papers. \toolname{} starts from a comparison query and retrieves numerical evidence from the open web. It uses a shared ontology to align records by entity, attribute, time, and scope. When reports still disagree, it shows the distribution of reported values and links them to their sources.}

\begin{figure*}
    \centering
    \includegraphics[width=\linewidth]{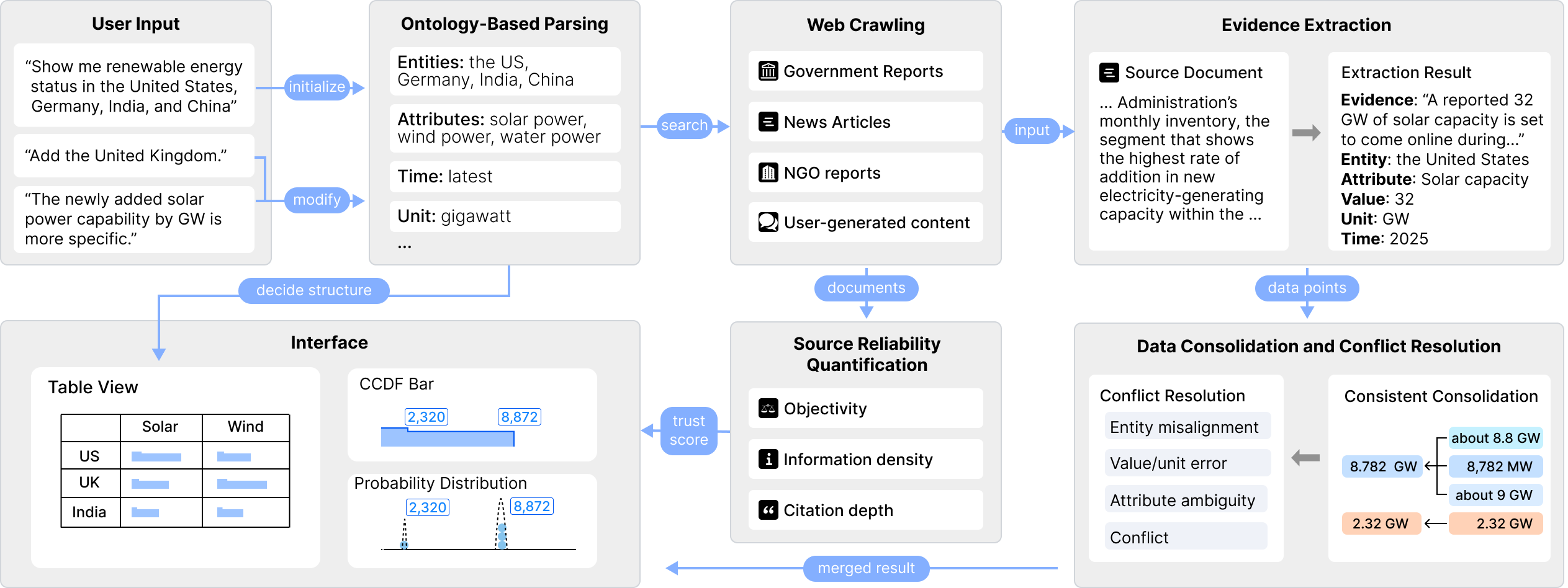}
    \caption{
    Overview of the \toolname{} workflow.
    A natural language input is parsed into a schema consisting of entities, attributes, time, and scope, which determines the table structure in the interface.
    The schema also guides web crawling and evidence extraction from diverse sources.
    Extracted evidence is then consolidated, assessed for source trustworthiness, and integrated into the final table and visualizations shown to the user.
    }
    \label{fig:pipeline}
\end{figure*}

\subsection{Data Trustworthiness}

\revision{Assessing conflicting web sources requires source reliability estimation and consistency validation~\cite{li2016survey}.}




\textbf{Source Reliability Estimation.}
This line of work assesses the credibility of information providers through both algorithmic modeling and human-centered analysis.
Automated approaches infer source reliability from signals, such as interaction patterns and propagation behaviors~\cite{castillo2011information}, and structural anomalies using graph-based techniques~\cite{Kaghazgaran2018TwoFace}, through fine-grained~\cite{ma2015faitcrowd} and bias-aware~\cite{li2020towards} probabilistic inference.
Complementarily, visual analytics research has framed trust as an interpretable and human-in-the-loop process, emphasizing data provenance~\cite{Ragan2016CPVDA}, transparency~\cite{7536106}, and interactive sensemaking~\cite{https://doi.org/10.1111/cgf.14034}. 
Building on these principles, several representative systems include those for the detection of anomalous user-generated content~\cite{cao2015targetvue, 6327271}, and the analysis of anomalous information spreading~\cite{zhao2014fluxflow}. 

\textbf{Consistency Validation.}
The second line of work establishes truth by validating extracted facts against logical constraints or prior knowledge. 
Existing methods enforce logical rules~\cite{ortona2018robust, ye2022constrained} or align extracted facts with external knowledge bases~\cite{dong2014knowledge} to detect inconsistencies. 
In parallel, visual analytics systems, such as QualityTrails~\cite{bors2014qualitytrails}, enable users to trace data transformations and assess consistency through transparent and interactive workflows.
\revision3{FOX~\cite{li2026fox} groups facts extracted from a dataset by analytical scope and helps users explore outliers within each group.}

Existing approaches often provide limited support for verifying open-domain web content.
\revision{Our framework addresses this by combining source reliability estimation and cross-source consistency inspection in a transparent process, helping users examine trust-related evidence more explicitly.}

\subsection{Uncertainty Visualization}


Communicating uncertainty is essential for reliable analytical decision-making, especially in settings where data are incomplete, noisy, or potentially conflicting~\cite{pang1997approaches, song2019where}.
Existing visualization approaches for uncertainty communication generally follow two paradigms: extrinsic auxiliary elements and intrinsic channel modifications~\cite{gershon1998visualization}.

\textbf{Extrinsic Techniques.}
Extrinsic methods add auxiliary graphical elements to depict distributions or error ranges without changing the original marks.
Common examples include error bars~\cite{correll2014error} and uncertainty glyphs~\cite{lu2020winglets, sanyal2009user, athawale2024uncertainty}.
For instance, Winglets~\cite{lu2020winglets} attach dual-sided strokes to scatterplot points, using their length and orientation to encode cluster association and uncertainty.
Such designs make uncertainty explicit and can improve perceptual grouping beyond color alone.

\textbf{Intrinsic Techniques and Integrations.}
Intrinsic methods instead modify existing visual channels, such as opacity and sketchiness~\cite{song2019where, 6327283}, to communicate confidence~\cite{MacEachren01012005, potter_quantification_2012}.
Other encodings, including value-suppressing palettes~\cite{10.1145/3173574.3174216}, Bayesian surprise maps~\cite{7536212}, and quantile dotplots~\cite{10.1145/2858036.2858558, 10.1145/3173574.3173718}, have been shown to reduce inference errors.
Recent systems further integrate uncertainty into analytical workflows, including feature attribution systems such as Explainer~\cite{8807299} and VBridge~\cite{9555810}, natural-language interfaces such as DataTone~\cite{10.1145/2807442.2807478} and Eviza~\cite{10.1145/2984511.2984588}, temporal ambiguity analysis~\cite{7192667}, and storytelling~\cite{8031599}.

Existing techniques provide limited support for representing conflicts across multiple sources.
To address this limitation, we propose a novel encoding that combines bars with cumulative distribution function curves to visualize uncertainty and data distribution.





\section{Overview}\label{sec:method}



As illustrated in \autoref{fig:pipeline}, \toolname{} consists of two major components: (1) a \textbf{trustworthiness-oriented data engine} (\autoref{sec:data_engine}) for evidence acquisition, organization, and reliability estimation, and (2) an \textbf{interactive interface} (\autoref{sec:interface}) for user-facing exploration and analysis.
Our system pipeline is designed to address the aforementioned four challenges.
To address incomplete retrieval results (\textbf{C1}), \toolname{} first transforms the users' natural-language input into a structured schema through \emph{ontology-based parsing}. 
This ontology-derived schema defines the initial layout of the output table and provides top-down guidance for subsequent retrieval. 
\revision{By explicitly specifying target entities and attributes, this schema-guided process directs retrieval toward individual table cells and makes under-supported cells explicit.}
To address sparse data availability across web sources (\textbf{C3}), \toolname{} then performs targeted \emph{web crawling} over diverse sources, such as government reports and news articles. 
The \emph{evidence extraction} module identifies relevant data points from these lengthy and disconnected documents, producing candidate evidence for subsequent integration. 
To address evidence conflicts from heterogeneous sources (\textbf{C4}), \toolname{} employs a \emph{data consolidation and conflict resolution} module. 
This stage aligns values under a unified schema, consolidates comparable evidence, and analyzes discrepancies caused by dimensional mismatches or factual conflicts. 
To support traceability throughout the process (\textbf{C2}), \toolname{} quantifies source reliability and presents the consolidated results in an interactive table view augmented with trust scores and provenance evidence. 
This design allows users to inspect underlying distributions, conflicts, and uncertainties, and to trace specific values back to their supporting documents.

\section{Trustworthiness-Oriented Data Engine}
\label{sec:data_engine}

The pipeline begins with \textbf{ontology-based parsing} (\autoref{subsec:ontology_parsing}), which extracts multidimensional coordinates, such as entity, attribute, time, and unit, from user input to define the fundamental structure of the analytical space.
Guided by these coordinates, the system performs \textbf{web crawling} and \textbf{evidence extraction} (\autoref{sec:evidence_extraction}) over diverse web sources to collect relevant raw facts.
To improve data robustness, the retrieved facts then undergo a dual-track refinement process. The \textbf{data consolidation and conflict resolution} module (\autoref{sec:conflict_resolution}) mitigates inconsistencies by aligning scales and resolving contextual mismatches, while the \textbf{source reliability quantification} module (\autoref{sec:source_reliability}) assesses the credibility of sources.


\begin{table}[h]
\centering
\caption{The 6-tuple ontology for data facts.}
\label{tab:ontology_tuple}
\resizebox{\linewidth}{!}{
\begin{tabular}{@{}lll@{}}
\toprule
\textbf{Component} & \textbf{Description} & \textbf{Example} \\ \midrule
Entity (\textbf{E}) & Subject serving as the comparison anchor. & ``Germany'' \\
Attribute (\textbf{A}) & Specific dimension or property being measured. & ``HSR Length'' \\
Time (\textbf{T}) & Temporal timestamp for trend analysis. & ``2020'', ``2023'' \\
Scope (\textbf{S}) & Shared contextual constraints that condition the fact. & ``Singapore market'' \\
Value (\textbf{V}) & Numerical value under comparison. & ``1500'' \\
Unit (\textbf{U}) & Measurement unit for numerical alignment. & ``km'', ``USD'' \\
\bottomrule
\end{tabular}%
}
\end{table}



\subsection{Ontology-Based Parsing}
\label{subsec:ontology_parsing}

\revision{To address contextual misalignment in unstructured web text, we define an ontology-inspired six-tuple schema that serves as a coordinate system for extracted facts.}
Each extracted fact is represented in six dimensions: Entity ($E$), Attribute ($A$), Time ($T$), Scope ($S$), Value ($V$), and Unit ($U$), as summarized in \autoref{tab:ontology_tuple}.
For example, from the sentence \textit{``In 2022, Toyota sold 2 million cars and BMW sold 1.5 million cars in Europe,''} we extract two distinct facts. Both share the same attribute (car sales), time (2022), and geographic scope (Europe), but they capture different entities and values: Toyota at 2 million units, and BMW at 1.5 million units.
Crucially, the assignment of these dimensions depends on context.
While ``Europe'' functions as a shared scope constraint in the previous example, in a query such as \textit{``Tesla sales in Germany and India,''} the countries instead become entities because they serve as the primary subjects of comparison.
This ontology directly addresses a root cause of data inconsistency.
Consider a query for the \textit{``annual car sales of Tesla in Germany,''} which might return two seemingly contradictory values: 200{,}000 and 250{,}000.
Without contextual awareness, a system might incorrectly identify them as a factual conflict.
However, our ontology-based coordinate space distinguishes these values by assigning them to different temporal contexts, for example, one to 2022 and the other to 2023.
Formally, because they differ along the time dimension, no true conflict exists.
By attaching contextual metadata to every data point, the schema transforms isolated values into context-bearing, inspectable records and enables users to align facts along shared dimensions, such as country, year, and metric, for presentation in a comparable tabular form.

\begin{figure}[ht]
    \centering
    \includegraphics[width=\linewidth]{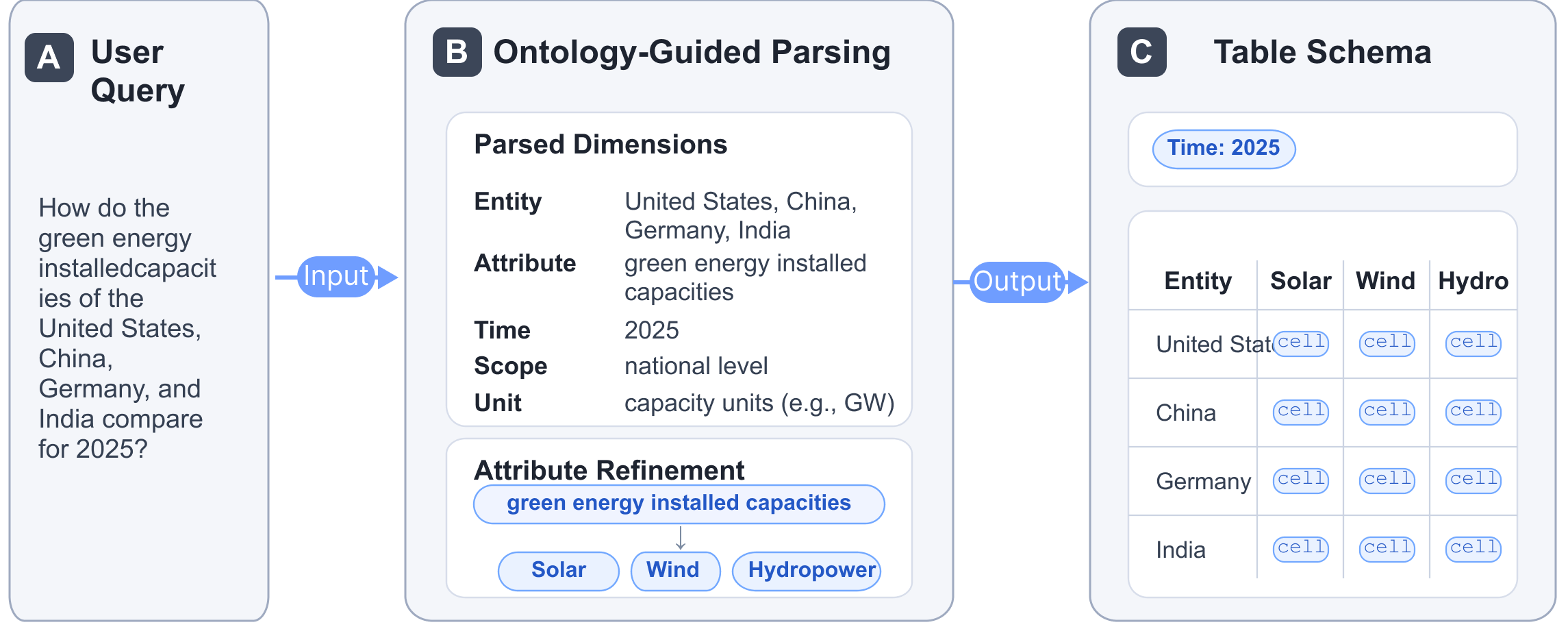}
    \caption{\revision{From natural-language query to table schema. 
\toolname{} parses a comparative query into entities, time, scope, and attribute intent, 
then expands the attribute into measurable dimensions to generate the target table schema.}}
    \label{fig:query_to_schema}
\end{figure}

\revision{Building upon this schema, our operational pipeline defines four transformation types that expand, align, and refine the analytical space.
\autoref{fig:query_to_schema} illustrates how ontology-guided parsing converts a comparative natural-language query into an explicit table schema. 
In this example, the system identifies the target entities and year, interprets the coarse attribute intent (``green energy installed capacities''), and refines it into measurable columns for subsequent retrieval.}
\revision{We operationalize this schema through four transformation types:}

\begin{itemize}
\item \textbf{Entity handling:}
Expands the comparison set with related entities to reduce sparsity and support cross-entity analysis.

\item \textbf{Attribute handling:}
Maps underspecified analytical intents to measurable attributes and adjusts attribute granularity when needed.

\item \textbf{Time handling:}
Extracts timestamps to align facts on a shared temporal axis and avoid conflating values from different years.

\item \textbf{Scope handling:}
Enforces scope matching over contextual constraints in $S$ so that only contextually aligned facts are compared.
\end{itemize}

These operations provide a common vocabulary for the transformations supported by the interface; user interactions are implemented through one or more of these operations.

\subsection{Web Crawling and Evidence Extraction}
\label{sec:evidence_extraction}

Based on the proposed ontology, we retrieve and extract evidence for each table cell defined by an entity, attribute, time, and scope.

\textbf{Web crawling and text collection.}
For each cell, we construct a query by combining the entity name, attribute keyword, temporal constraint, and, when applicable, the scope, such as a country or region.
\revision{We send each query to Google via the Serper API\footnote{\url{https://serper.dev/}} and retrieve up to 20 ranked results by default. \revision3{The limit is configurable.}}
We download the returned pages and clean them by removing HTML tags, navigation menus, and advertisements.
We then extract the cleaned text body and remove duplicate \revision3{paragraphs} to avoid repeated content~\cite{dong2014knowledge}.
The resulting corpus spans diverse sources, including government reports, news articles, NGO documents, independent media reports, and personal blogs.



\textbf{LLM-Assisted Evidence Extraction.}
We use an LLM-based pipeline to transform unstructured text into ontology-consistent structured evidence.
Specifically, we prompt the LLM to extract relevant evidence for each target cell and return the result in a structured format. The prompt includes the field definitions derived from the ontology, together with an illustrative example, to guide the model toward consistent outputs.
For each text segment and target cell, defined by scope, entity, attribute, and time, the LLM returns one or more records containing the entity, attribute, time, scope, value, unit, a relevance score, and the supporting sentence(s).
For example, suppose the target cell corresponds to \textit{entity = Germany}, \textit{attribute = solar power total installed capacity}, and \textit{time = 2025}.
If a retrieved web page contains a sentence such as ``At the end of 2025, the total installed solar capacity in Germany reached 117 GW,'' the LLM extracts a record with \texttt{value = 117}, \texttt{unit = GW}, and the quoted sentence as supporting evidence.
These supporting sentences serve as human-readable justifications, while the structured fields are used for downstream aggregation and visualization.
Multiple records may correspond to the same cell.
We retain all such records for subsequent deduplication, conflict resolution, and confidence estimation.


\subsection{Data Consolidation and Conflict Resolution}
\label{sec:conflict_resolution}

The data merging process is conducted in two sequential steps: interval-based data consolidation and LLM-assisted conflict resolution.

\textbf{Interval-Based Data Consolidation.}
The first step aims to merge non-conflicting values extracted from different sources whenever possible. Such cases often arise when sources report values at different granularities or in different interval forms. For example, Source A may report a value larger than three, Source B may report $3.5$, and Source C may report $3.54$. \revision3{The intervals implied by the three reports overlap, so we treat the reports as compatible.}
To assess this consistency, we convert the reported values into intervals and examine their overlaps. Numerical precision can implicitly convey uncertainty. When uncertainty is not explicitly reported, the last significant digit implies an interval of approximately $\pm 0.5$ at that level of precision.
For example, $3.78$ implies an interval of $[3.775, 3.785]$~\cite{iso1993guide}.
By representing values as intervals, we can identify compatible reports and consolidate them accordingly.
\revision{For a group of mutually consistent values, we use the highest-precision value as a display representative while retaining all supporting records for inspection.
This choice preserves the finest reported granularity but does not treat numerical precision itself as evidence of factual accuracy.}
After this step, multiple distinct results may still remain, and we therefore further resolve these apparent discrepancies through conflict resolution, as their underlying causes may differ.


\textbf{LLM-Assisted Conflict Resolution.}
After the consolidation phase, unresolved discrepancies may still exist.
To address them, we introduce an LLM-assisted approach to conflict resolution.
This module examines the original textual evidence, the extracted values, and the conflicting results to diagnose the underlying cause of each discrepancy.
We categorize the causes of conflicts into the following four types and apply tailored resolution strategies:
\begin{itemize}
\item \textbf{Entity misalignment:}
The extraction incorrectly targets the wrong entity or scope (e.g., extracting global values when U.S. values are required). \textit{Resolution:} Data points identified with entity misalignment are filtered out.
\item \textbf{Value or unit extraction errors:}
These are system-level failures during the NLP extraction process, including incorrect numeric parsing or unit conversion errors. \textit{Resolution:} Upon detecting such errors, the system automatically triggers a re-extraction process for the affected data points.
\item \textbf{Attribute semantic ambiguity:}
The extraction is technically correct, but the queried attribute is inherently ambiguous and requires further specification (e.g., ``solar capacity" could mean newly installed capacity'' or ``cumulative capacity'').
\textit{Resolution:} The system defers to the user by prompting them in the chat view (\autoref{sec:chat_view}).
The user can either specify a more precise definition or choose to retain both interpretations as newly defined attributes.
\item \textbf{Genuine Factual Contradiction:}
\revision{The extracted records appear consistent with their source texts, but the underlying sources report incompatible values.}
Because such conflicts cannot be resolved automatically without additional domain evidence, the system retains the competing records and summarizes their relative source-supported weights using the source-quality scores (detailed in \autoref{sec:source_reliability}).
The distribution and the conflicting sources are then exposed in the user interface, empowering users to make the final judgment based on their domain knowledge.
\end{itemize}

\begin{figure}
    \centering
    \includegraphics[width=\linewidth]{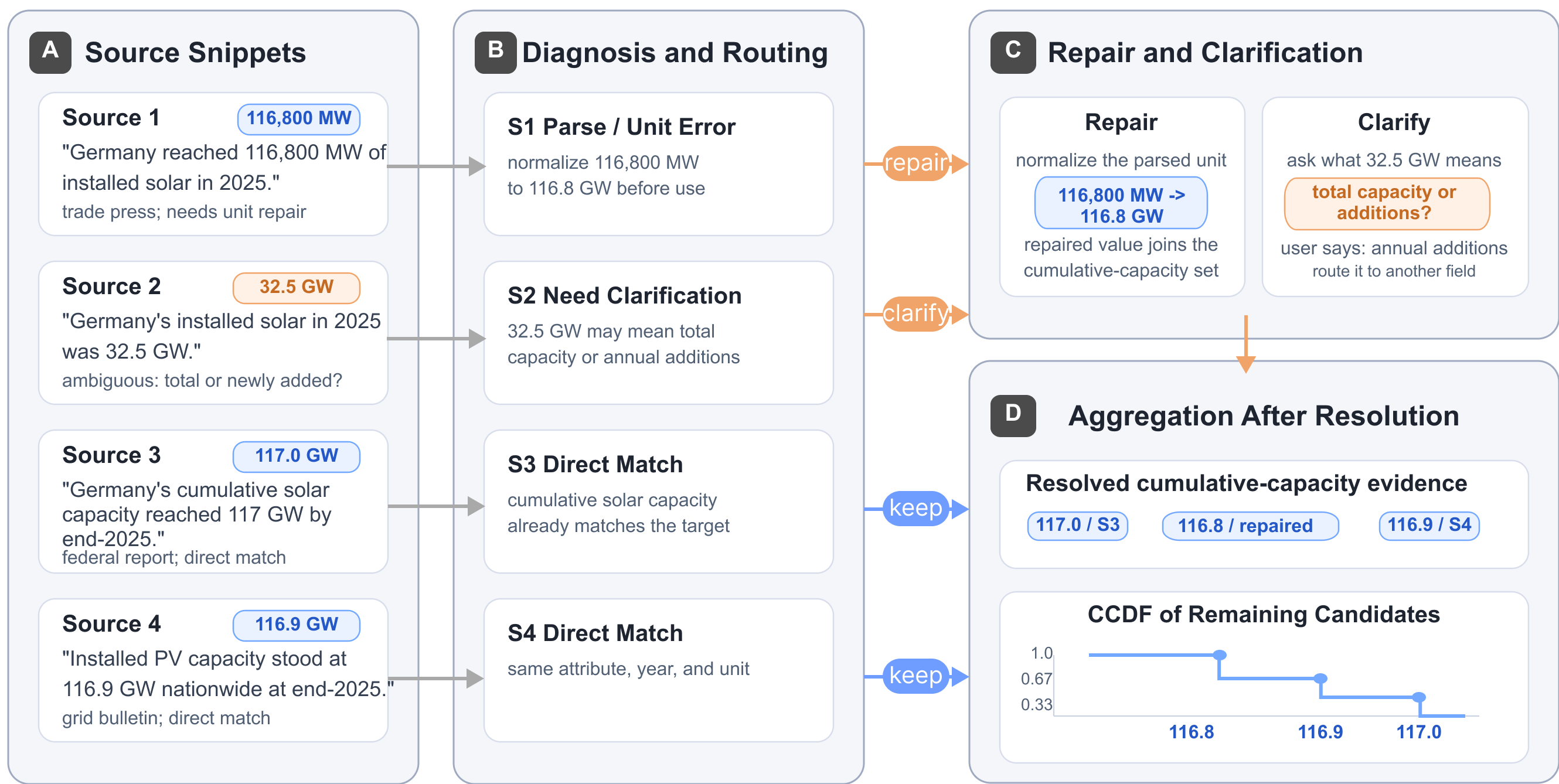}
    \caption{Conflict resolution via source cases.}
    \label{fig:conflict_aggregation_example}
\end{figure}

\revision{\autoref{fig:conflict_aggregation_example} illustrates the conflict-resolution workflow with four source cases.
The reports that already match the target attribute proceed directly to aggregation, whereas problematic cases are routed to either repair or clarification.
In the example, a value reported as 116,800 MW is normalized to 116.8 GW and then re-enters the cumulative-capacity evidence pool, while an ambiguous 32.5 GW report is clarified as annual additions and therefore rerouted to a different attribute.
Only resolved cumulative-capacity candidates are passed to the final CCDF encoding.}

\subsection{Source Reliability Quantification}
\label{sec:source_reliability}

We define the trust score \(T(s)\) as a normalized, heuristic source-quality score for a document \(s\). 
It combines citation evidence, information density, and stylistic neutrality to prioritize sources within the retrieved evidence set; it is not intended as a calibrated probability that an individual claim is correct.

\begin{itemize}
\item \textbf{Citation Depth ($v_{cite}$).}
Motivated by prior work on knowledge-based trust~\cite{dong2015knowledge}, we use the presence of traceable citations as one inspectable proxy for source quality.
We quantify citation depth using a hyperbolic tangent function to normalize the count of references as \revision{$$v_{cite} = \tanh((N_{ext} + \kappa \cdot N_{acad}) / \tau),$$}
where $N_{ext}$ represents the count of valid external hyperlinks, and $N_{acad}$ denotes the count of references adhering to academic standards.
The parameter $\kappa > 1$ weights academic citations higher than general links, and $\tau$ is a saturation constant controlling the marginal utility of additional citations.
Empirically, we set $\kappa = 2$ to emphasize scholarly rigor and $\tau = 5$ to model the principle of diminishing returns (e.g., 5 weighted citations yield $\tanh(1) \approx 0.76$).

\item \textbf{Information Density ($v_{dens}$).}
To measure the substantive content ratio (i.e., signal-to-noise ratio) of the text, we calculate the density of named entities and numerical data: \revision{$$v_{dens} = \sum_{t \in T} (\mathbb{I}(t \in \mathcal{E}_{NER}) + \mathbb{I}(t \in \mathcal{N}_{num})) / |T|,$$}
where $T$ is the set of valid tokens (excluding stop words) in the text, and $\mathbb{I}(\cdot)$ is the indicator function. $\mathcal{E}_{NER}$ and $\mathcal{N}_{num}$ represent the sets of named entities and numerical tokens.

\item \textbf{Objectivity ($v_{obj}$).}
Following the definition of data objectivity \cite{wang1996beyond}, we penalize sources exhibiting strong emotional polarity, which often correlates with subjective bias. The objectivity score is defined as $v_{obj} = 1 - | \text{Polarity}(T) |^\gamma$, where $\text{Polarity}(T) \in [-1, 1]$ is the output of a pre-trained sentiment analysis model (i.e., BERT-Sentiment).
We assign $\gamma = 1.5$ to exponentially penalize extreme sentiment polarity while preserving subtle emotional expressions.

\end{itemize}


The final trust score is computed using a logistic function $T(s)=\sigma(\mathbf{w}^T \mathbf{v}_s + b)$, where $\sigma(\cdot)$ is the sigmoid function mapping the result to $(0,1)$.
\revision3{We use heuristic defaults of $\mathbf{w}=[1.2, 3.0, 0.6]^T$ and $b=-1.2$, which users can adjust in the configuration panel.
When we varied one weight at a time by $\pm30\%$, all Spearman rank correlations with the default ranking exceeded $0.98$.}

\begin{figure}
    \centering
    \includegraphics[width=\linewidth]{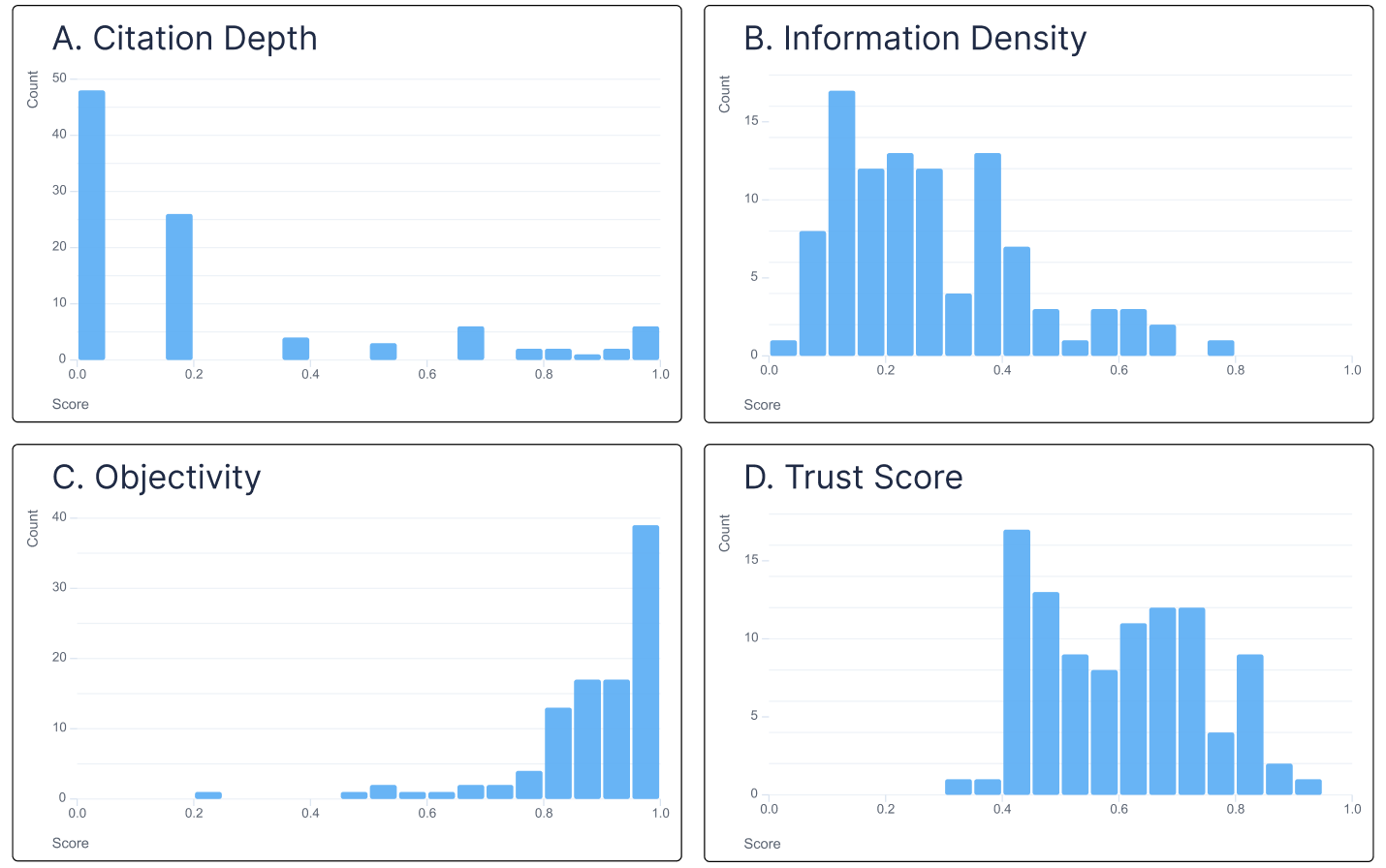}
    \caption{The distributions of the three components of our source reliability model and the final trust score, computed on 100 web documents.}
    \label{fig:reliability_distribution}
\end{figure}

\textbf{Distribution of the Trust Scores.}
\autoref{fig:reliability_distribution} shows the distributions of the three components and the final trust scores for 100 web documents sampled from our corpus. Citation depth is highly skewed, information density varies more broadly, and the final scores span a broad range.

\section{Interface of \toolname{}}
\label{sec:interface}
The interface of \toolname{} is designed to support seamless transitions between high-level exploratory analysis and low-level data verification. It is organized into four coordinated components: the traceability inspector (\autoref{fig:interface}~\captionID{A}) for detailed provenance inspection, the metadata panel (\autoref{fig:interface}~\captionID{B}) for schema management, the main table view (\autoref{fig:interface}~\captionID{C}) for result presentation, and the chat view (\autoref{fig:interface}~\captionID{D}) for intent-driven interaction.


\subsection{Chat View and Metadata Panel}
\label{sec:chat_view}

The chat view (\autoref{fig:interface}~\captionID{D}) serves as the natural-language entry point to \toolname{}.
It supports several key interaction capabilities:

\begin{itemize}
    \item \textbf{Initializing analysis intent.}
    Users can issue high-level, fuzzy queries (e.g., ``Compare green energy capacities''), which the system automatically parses to construct an initial table schema with appropriate entities and attributes.

    \item \textbf{Iterative schema refinement.}
    Users can refine the table by adding or modifying entities, attributes, or scopes through natural-language commands.

    \item \textbf{Human-in-the-loop semantic clarification.}
    When attributes are ambiguous or overly coarse, users can negotiate more precise definitions with the system. In response, the system suggests refined schema variants and updates the table accordingly.

    \item \textbf{Proactive ambiguity and conflict surfacing.}
    When the background data engine detects semantic ambiguities or conflicting definitions (e.g., inconsistent interpretations of a single attribute), the assistant proactively raises an ``ambiguity detected'' alert and suggests schema refinements, such as splitting or redefining attributes.
\end{itemize}

The parsed results of user queries are explicitly displayed in the metadata panel (\autoref{fig:interface}~\captionID{B}), which is located above the table.
This panel organizes the analysis into three structured facets: \textit{Scope and Time}, \textit{Entities} (e.g., countries), and \textit{Attributes} (e.g., solar power capacity).
Users can directly intervene in this structured representation, for example, by clicking the ``+ ADD'' button to introduce new entities or refine contextual constraints, ensuring that the system's retrieval boundaries remain aligned with their analytical intent.


\begin{figure}[ht]
    \centering
    \includegraphics[width=\columnwidth]{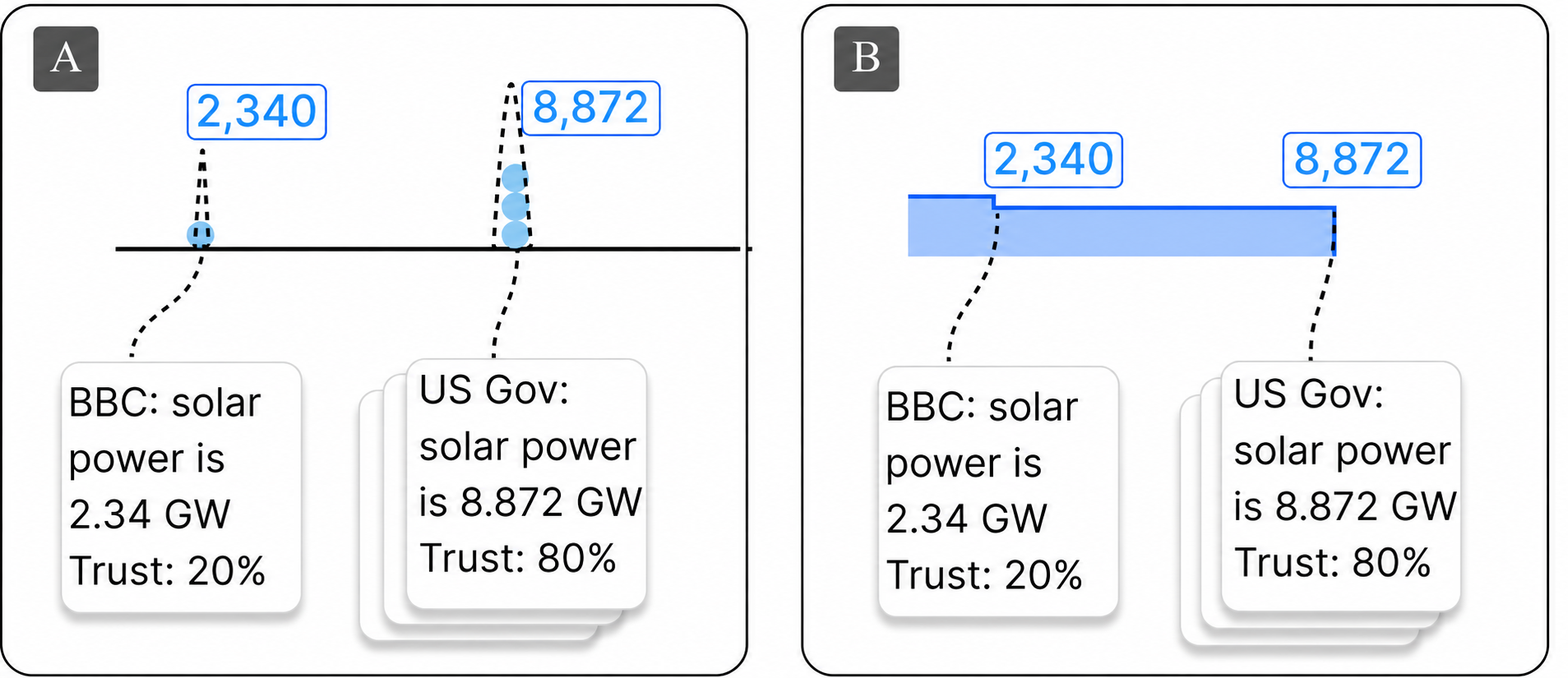}
    \caption{Two visual encodings for visualizing contradictions in source data. \revision3{Plot labels use MW, while source cards report GW ($1\,\mathrm{GW}=1{,}000\,\mathrm{MW}$). The curves are schematic; mixtures of uniform interval densities yield piecewise-constant densities and piecewise-linear CCDFs.}}
    \label{fig:uncertainty}
\end{figure}

\subsection{Table View}
\label{sec:table_view}

\begin{figure*}[ht]
    \centering
    \includegraphics[width=\textwidth]{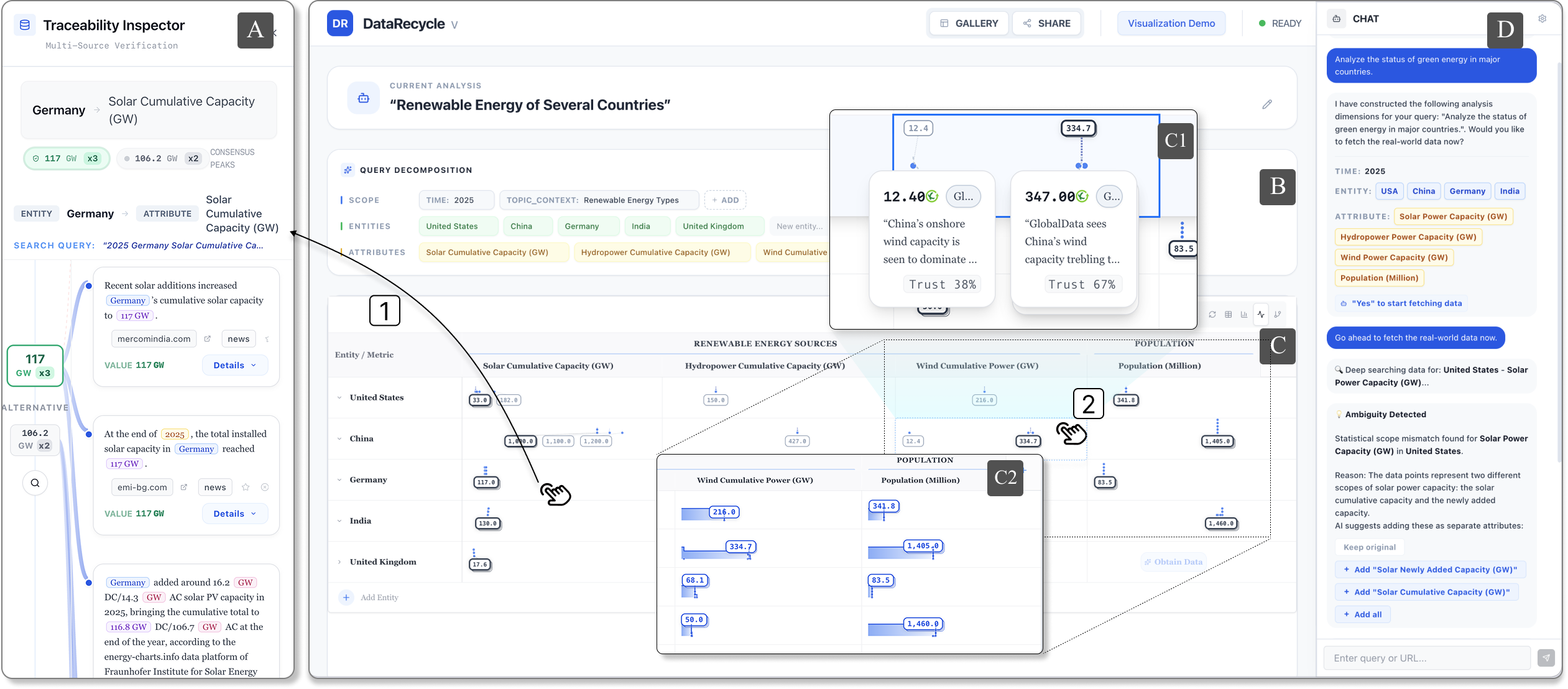}

    \caption{The user interface of \toolname{}. (A) A provenance inspector that reveals the specific data source of a cell's value upon clicking. (B) A metadata panel displaying the entities, attributes, and scope of the current table. (C) The main table view presenting the data results, equipped with a lens tool (C1) that allows users to hover and examine detailed information. (D) A chat interface for users to input natural language queries and interact with the system.}

    \label{fig:interface}

\end{figure*}

The table view (\autoref{fig:interface}~\captionID{C}) serves as the primary workspace for data presentation.
Each row represents an entity, and each column represents an attribute.
To handle the inherent uncertainty and multiplicity of web-extracted data, the table provides two specialized visual encodings, as shown in \autoref{fig:uncertainty}.

\begin{itemize}
\item \textbf{Probability distribution encoding.}
For each cell, we visualize multiple extracted values collected from different sources. \revision3{We build the density from the original records with finite intervals and use one-sided bounds only to check compatibility.} Each \revision3{eligible} extracted value $x_i$ is modeled as a \revision3{uniform local density} $p_i(x)$ over an interval $I_i = [x_i - \delta_i, x_i + \delta_i]$, where \revision3{the half-width $\delta_i>0$} is determined by the reported precision of $x_i$. The density satisfies $p_i(x) \geq 0$ and $\int p_i(x)\,dx = 1$. \revision3{Specifically, $p_i(x)=1/(2\delta_i)$ for $x\in I_i$ and $p_i(x)=0$ otherwise.} 
Given trust scores $\{t_i\}_{i=1}^{n}$, we aggregate these local densities into a normalized, source-weighted evidence-support distribution, $P(x) = \sum_{i=1}^{n} \alpha_i p_i(x)$, where $\alpha_i = t_i / \sum_{j=1}^{n} t_j$ and $\sum_{i=1}^{n}\alpha_i = 1$. Thus, sources with higher trust scores contribute more strongly to the final distribution. \revision3{The evidence-support model uses this mixture directly, without kernel smoothing. Its density is piecewise constant, and its CCDF is continuous and piecewise linear.} In \autoref{fig:uncertainty}~\captionID{A}, each source is represented by a small dot positioned according to its extracted value along a horizontal value axis shared by all cells in the same attribute column. To reduce overplotting, nearby dots are vertically jittered or stacked. The aggregated distribution $P(x)$ is shown as an overlaid density curve. This encoding reveals variance, consensus peaks, and outliers, allowing users to assess how concentrated or dispersed the evidence is for a given metric. As illustrated in \autoref{fig:uncertainty}, this example contains four sources. Three sources report the value \revision3{8.872 GW}, each with a trust score of 80\% (computed as described in \autoref{sec:source_reliability}), whereas one source reports \revision3{2.34 GW} with a trust score of 20\%. Accordingly, the density curve places greater emphasis on the cluster around \revision3{8.872 GW} while still preserving the lower-confidence evidence at \revision3{2.34 GW}.

\item \textbf{CCDF bar encoding.}
Although the probability-distribution encoding is intuitive for a single cell, it leaves much of the cell area unused when applied to large tables.
To address this limitation, we introduce a more space-efficient visualization based on a CCDF bar (\autoref{fig:uncertainty}~\captionID{B}) to represent the same uncertainty information.
When users need to compare magnitudes across entities while retaining uncertainty context, they can switch to the CCDF bar encoding (\autoref{fig:interface}~\captionID{C2}).
We introduce a bar-like visual encoding that represents the remaining evidence-support mass along the value axis.
Let $X$ \revision3{follow the aggregated evidence-support distribution}, and let $F(x)$ denote its cumulative distribution function (CDF). We define the \revision3{CCDF} as: $R(x) = P(X > x) = 1 - F(x)$.
The CCDF bar visualizes $R(x)$ as the profile of a horizontal bar over the value axis.
The bar begins at full height (probability $1$) and progressively decreases as probability mass accumulates along the axis.
For discrete data points clustered from web sources with value probability pairs $D = \{(v_1,p_1), \dots, (v_n,p_n)\}$, this yields a descending staircase structure:
$R(x) = 1 - \sum_{v_i \le x} p_i$.
For continuous distributions with density $f(x)$, it forms a \revision3{continuous, non-increasing profile}:
$R(x) = \int_x^{\infty} f(t)\,dt.$
\revision3{Bar extent marks the upper end of the support. To compare entities, read $R(x)$ at the same threshold $x$; extent alone gives no general probability ordering. $R(x)$ describes the evidence-support model, not a calibrated probability of truth.}

\end{itemize}

\begin{figure}[t]
    \centering
    \includegraphics[width=\columnwidth]{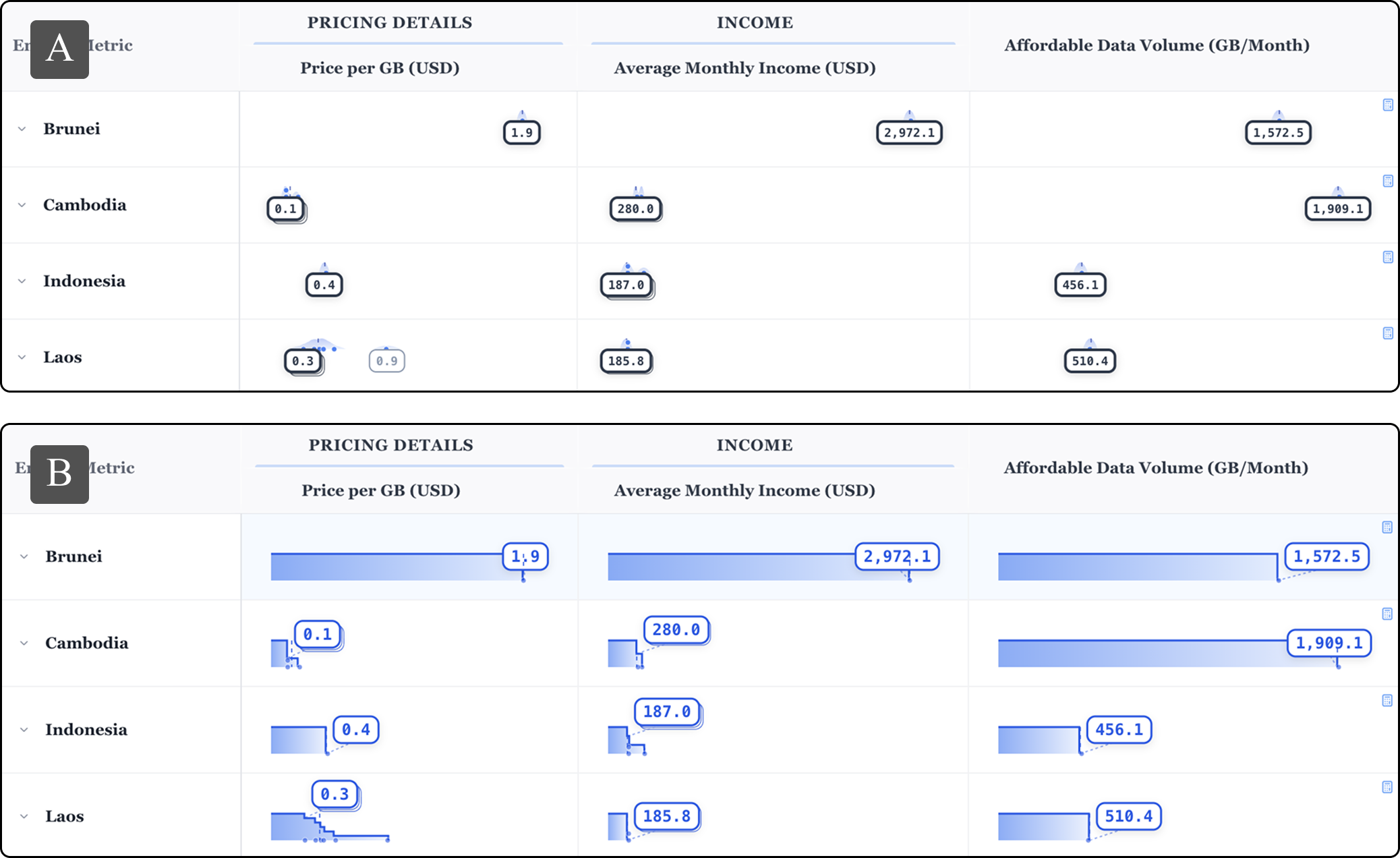}
    \caption{Cross-entity comparison with (A) probability distributions and (B) CCDF bars. \revision3{Each CCDF bar ends at the distribution's upper bound. Its height at each threshold shows the share of evidence weight above it.}}
    \label{fig:table_compare}
\end{figure}

\noindent As shown in \autoref{fig:table_compare}, \captionID{A} presents the probability distribution encoding, and \captionID{B} shows the CCDF encoding.
Both encodings compare mobile data prices and average monthly income across several countries.
The CCDF encoding can be interpreted as an enhanced bar chart: when the values reported by different sources are highly consistent, it nearly collapses to a conventional bar chart.
\revision3{For the price per GB in Laos, the profile falls over different price ranges. The horizontal gaps between these declines show how far apart the reported prices are.}
Compared with the probability-distribution view, the CCDF view is more compact and supports rapid scanning across rows and columns.
Beyond visual encodings, the table view also supports interactive operations that enable users to inspect uncertainty and derive new insights from the data.

\begin{itemize}

\item \textbf{Semantic Lens.}
Due to limited cell space, displaying the full context of conflicting values simultaneously is impractical.
To address this issue, we designed a semantic lens tool.
When a user hovers over a cell, the lens expands to juxtapose the distinct values.
As shown in \autoref{fig:interface}~\captionID{C1}, the lens provides a set of summary cards for each extracted value.
Each card displays the numerical value, the supporting extraction context, and a calculated trust score for the source.
This design allows users to inspect conflicting data in situ without losing the surrounding table context.

\item \textbf{Attribute Derivation.}
Users can click the attribute derivation button to create a new attribute based on expressions over existing attributes.
For example, in \autoref{fig:table_compare}, after obtaining the data price per GB and the average monthly income, we can derive a new attribute, ``affordable data volume'', defined as the ratio of income to data price.
This derived attribute enables more meaningful comparisons and provides an intuitive understanding of data affordability across countries.
\revision3{This calculation divides representative income by representative data price. The result is a point estimate without joint uncertainty propagation.}

\end{itemize}

\subsection{Traceability Inspector Panel}
The Traceability Inspector (\autoref{fig:interface}~\captionID{A}) is a collapsible drawer that exposes provenance details for the selected cell.
When a user selects a specific data cell in the table, the corresponding traceability panel is automatically revealed.
At the top of the inspector, we present an overview of the consensus peaks, summarizing the aggregated evidence for the selected cell.
Beneath this overview, the panel lists a set of evidence cards, each corresponding to an individual source that contributes to the selected cell.
Each evidence card exposes rich data lineage, including the source URL and its domain type (e.g., news outlet, government report), the extracted semantic tags (entity, attribute, time, and unit), and the exact text span from the original webpage that serves as evidence.
Users can interact with these cards to open the original source in an external browser, assess the quality of the automatic extraction, or explicitly exclude erroneous or low-quality sources from the final aggregation.

\section{Evaluation}
\label{sec:experiments}


\revision3{Our evaluation comprises four parts: a usage scenario illustrating the workflow, module benchmarks of evidence extraction and record-level auditing, a controlled perceptual comparison of the two encodings, and a system user study collecting usability and perceived-usefulness feedback.}

\subsection{Usage Scenario}
\label{sec:usage}


Using the green energy sector as an example, relevant data are typically distributed across websites with inconsistent statistical criteria.
Traditional search engines often return noisy or localized results.

\textbf{From Fuzzy Query to Initial Table and Metadata.}
Our approach allows users to begin exploration with a fuzzy natural language query.
For example, a user may ask in the chat view (\autoref{fig:interface}~\captionID{D}), ``How do the green energy installed capacities of the United States, China, Germany, and India compare for 2025?'' The system semantically parses the query and extracts the corresponding metadata, including Scope (Time), Entities, and Attributes (e.g., Solar, Hydropower, and Wind).
The extracted metadata are structured and displayed in the metadata panel (\autoref{fig:interface}~\captionID{B}).
Based on these dimensions, the system automatically generates an initial two-dimensional table view (\autoref{fig:interface}~\captionID{C}).
At this stage, the user can further refine the table, for example, by adding the UK as a new entity through either the table view or the chat view.


\textbf{Human-in-the-loop Conflict Resolution.}
After confirming the table structure, the user prompts the system to fetch data. The backend crawler then performs cell-level retrieval. However, open-web data are often contradictory because of mismatched entity granularity, differences in temporal scope, or attribute ambiguity. To address these issues, the Audit Agent performs preliminary cleaning and supports a human-in-the-loop workflow. As shown in the chat view (\autoref{fig:interface}~\captionID{D}), when the backend detects semantic ambiguity in an attribute, for example, when ``solar power capacity'' may refer to cumulative capacity or newly added capacity, it issues an ``Ambiguity Detected'' prompt. The system then suggests either retaining the original attribute or splitting it into more precise columns along additional dimensions, allowing the user to actively guide the data structuring process.


\textbf{Visual Exploration and Lens-based Inspection.}
Once the table is populated, the system provides rich visual analytics support.
By default, the table presents data as the probability distribution encoding to reflect variance or uncertainty.
To facilitate intuitive high-level comparisons across countries, the user can switch to the CCDF bar encoding (\autoref{fig:interface}~\captionID{C2}), which visually encodes data magnitude, for example, when comparing population or wind cumulative power across entities.
Furthermore, for cells containing complex or conflicting values, the user can activate the semantic lens (\autoref{fig:interface}~\captionID{C1}).
Hovering over or focusing on a specific area opens a detailed lens that juxtaposes distinct extracted values (e.g., 12.40 vs. 347.00) with their corresponding trust scores (computed as described in \autoref{sec:source_reliability}) and evidence sentences, helping users quickly assess data reliability.


\textbf{Multi-Source Traceability and Verification.}
Ensuring data accuracy is critical in policy research.
When a user wants to audit a specific cell's value (e.g., Germany's solar cumulative capacity of 117~GW), they can click on the corresponding data node (as indicated by \captionNum{1}).
This action expands the traceability inspector (\autoref{fig:interface} \captionID{A}) on the left.
This panel visualizes the ``Consensus Peaks'' of the extracted values (e.g., clusters around 117~GW and 106.2~GW). Crucially, it lists the exact provenance for each data point, including source URLs (e.g., \textit{emi-bg.com}), the original textual context, and the extracted time and attribute tags.
This transparency allows researchers to verify and select data sources for further analysis.



\begin{table}[ht]
    \centering
\caption{\revision{Performance of different LLMs on evidence extraction and \revision3{record-level evidence auditing}.}}
    \label{tab:comprehensive_evaluation}

    \setlength{\tabcolsep}{3pt}
    \resizebox{\columnwidth}{!}{
    \revisionblock{
    \renewcommand{\arraystretch}{1.1}
    \begin{tabular}{@{} l l ccc cc @{}}
        \toprule
        \multirow{2}{*}{\textbf{Family}} &
        \multirow{2}{*}{\textbf{Model}} &
        \multicolumn{3}{c}{\textbf{Evidence Extraction}} &
        \multicolumn{2}{c}{\textbf{\revision3{Evidence Auditing}}} \\
        \cmidrule(lr){3-5} \cmidrule(l){6-7}
        & &
        \makecell[c]{\textbf{Value} \\ \textbf{Acc. (\%)}} &
        \makecell[c]{\textbf{Unit} \\ \textbf{Acc. (\%)}} &
        \makecell[c]{\textbf{Evid.} \\ \textbf{Acc. (\%)}} &
        \makecell[c]{\textbf{Audit} \\ \textbf{Acc. (\%)}} &
        \makecell[c]{\textbf{Avg. Time} \\ \textbf{(s)}} \\
        \midrule

        GPT & 4o-mini      & 92.50 & 92.00 & 89.50 & 91.04 & 3.69 \\
        GPT & 4o           & 94.00 & 89.50 & 90.00 & 96.89 & 1.99 \\
        GPT & 5.4-mini     & 92.50 & \highlight{97.00} & 92.50 & 93.24 & \highlight{1.66} \\
        GPT & 5.4          & 97.50 & 91.00 & \highlight{97.50} & 96.34 & 2.23 \\
        \midrule
        Claude & Sonnet-5 & \highlight{99.00} & 95.50 & 95.00 & 95.25 & 5.98 \\

        \midrule
        Gemini & 3.1-Pro-Preview & 95.00 & 88.00 & 89.50 & \highlight{98.17} & 9.78 \\
        Gemini & 3-Flash-Preview & 97.50 & 90.00 & 94.50 & 96.53 & 3.61 \\
        Gemini & 3.1-Flash-Lite & 95.00 & 91.50 & 89.00 & 97.07 & 3.48 \\

        \midrule
        DeepSeek & v4-pro & 94.00 & 93.00 & 96.00 & 97.44 & 8.74 \\
        DeepSeek & v4-flash & 95.00 & 93.50 & 94.00 & 97.07 & 28.52 \\

        \midrule
        Doubao & seed-1-6-251015 & 80.50 & 84.00 & 85.50 & 97.99 & 34.74 \\
        \bottomrule
    \end{tabular}
    }}
\end{table}

\subsection{Quantitative Evaluation}
\label{sec:model_eval}

We evaluate two components of the data engine (\autoref{sec:data_engine}): evidence extraction and record-level auditing within the conflict-resolution stage.
\revision{The results are summarized in \autoref{tab:comprehensive_evaluation}.
\revision3{The extraction benchmark comprises 200 source--target cases from 186 cleaned source files across 12 attributes. Each provides a target, relevant text, and a reference value, unit, and evidence sentence.
Cases were selected with seed 203, using a confidence-based heuristic and query/source diversity.
Input lengths are 991--4,776 characters (median 3,785). The inputs are evidence windows rather than full retrieved pages.}
Given the query target and source text, each LLM was prompted to return a structured \texttt{JSON} output containing the extracted value, unit, and evidence sentence. 
We compared the outputs with the human annotations and report value, unit, and evidence accuracy under \textit{Evidence Extraction}. 
For conflict resolution, each benchmark case contains an expected query target and a list of candidate extracted records, each with an extracted entity, attribute, value, unit, and textual evidence. 
Following the conflict types defined in \autoref{sec:conflict_resolution}, each model was asked to assign one audit label to each candidate record. 
We evaluated this module on 200 audit cases containing 547 \revision3{records, including 200 with injected errors (Supplement A.4)}.
A prediction was counted as correct when the assigned label matched the human-annotated cause of the discrepancy.
\revision3{Three experts cross-annotated the records, resolved disagreements through discussion, and agreed on acceptable labels before inspecting model outputs. When two diagnoses were acceptable, either counted as correct.}
We report item-level audit accuracy and average inference time under \revision3{\textit{Evidence Auditing}}.
\revision3{The error labels correspond to entity/scope misalignment, value/unit extraction errors, and attribute ambiguity in \autoref{sec:conflict_resolution}.
The label \texttt{extraction\_correct} means that a record matches its source evidence and target context, even when sources disagree. The reported accuracy measures record-level auditing.}}

\textbf{Results.}
Model rankings differed between evidence extraction and record-level auditing (\autoref{tab:comprehensive_evaluation}). Gemini 3.1-Pro-Preview led auditing (98.17\%). Doubao seed-1-6-251015 scored 80.50\% on value extraction and 97.99\% on auditing.

\begin{figure}
    \centering
    \includegraphics[width=\columnwidth]{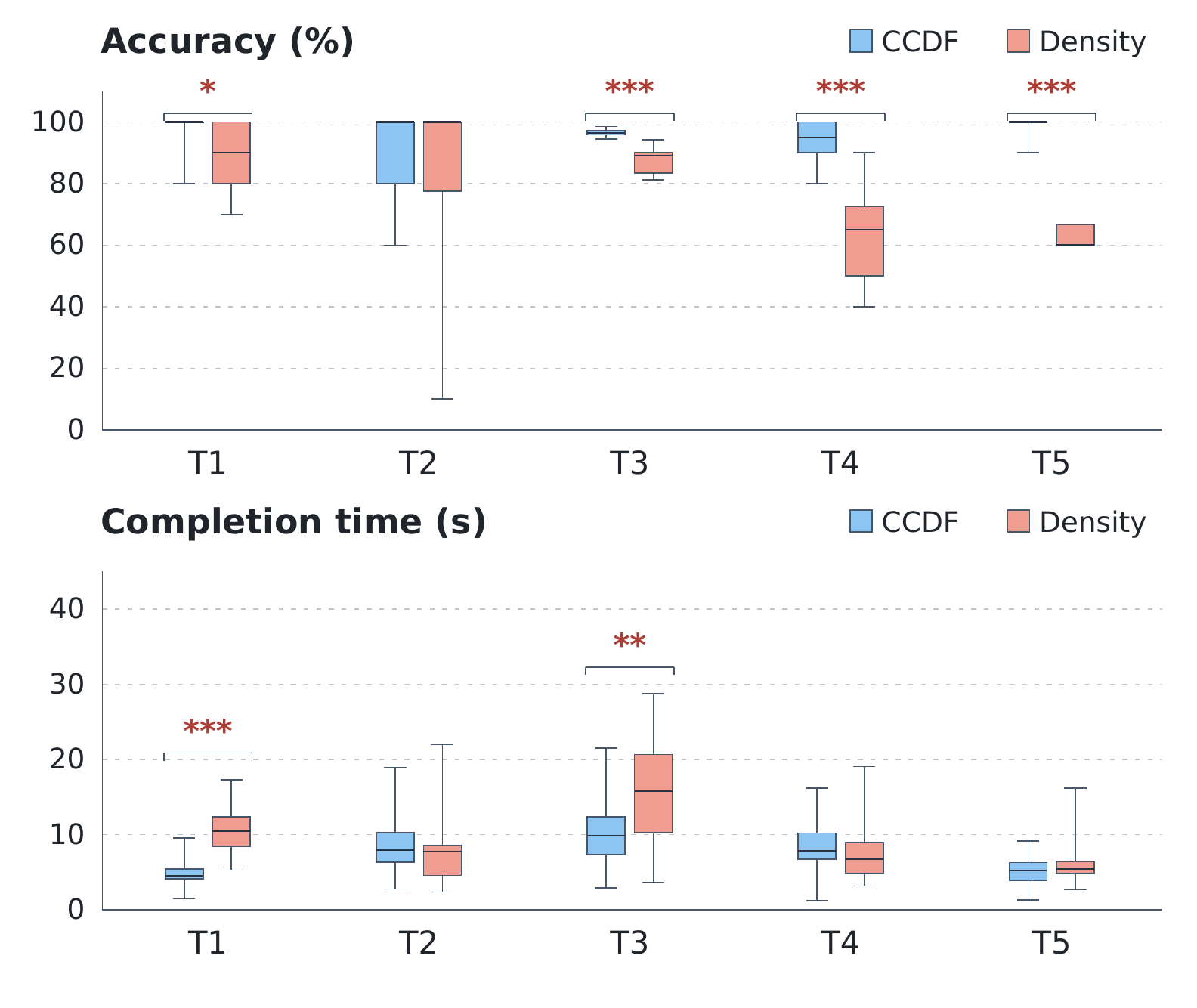}
    \caption{\revision{Controlled user study results comparing the CCDF-based bar encoding with the density-plot encoding on five uncertainty interpretation tasks. \revision3{Boxes summarize participant-level means: the box spans the 25th--75th percentiles, the central line is the median, and whiskers mark the minimum and maximum.} Significance markers use \(*\) for \(p<.05\), \revision3{\(**\) for \(p<.01\)}, and \(***\) for \(p<.001\).}}
    \label{fig:perceptual_study}
\end{figure}

\subsection{Perceptual Comparison}


\revision{We conducted a controlled user study comparing our CCDF-based bar encoding with a density-plot encoding (i.e., the probability distribution encoding described in \autoref{sec:table_view}) on uncertainty interpretation tasks.}

\revision{\textbf{Task Design.}
We designed five tasks to cover the core operations involved in analyzing uncertain distributions. T1 (\textit{Extremes}) asks participants to quickly identify the cell with the maximum or minimum value in a dense \(5 \times 5\) matrix, evaluating how well an encoding supports rapid visual scanning. T2 (\textit{Consistency}) asks participants to identify the most consistent and the least consistent cells in the matrix. T3 (\textit{Probability Estimation}) asks participants to estimate the probability that a value exceeds a given threshold. T4 (\textit{Cluster Weight}) asks participants to \revision3{select the cluster with the largest probability mass. The component weights determine the reference answer.} T5 (\textit{Entity Comparison}) asks participants to \revision3{identify the entity with the highest probability of exceeding a given threshold. Reference answers are computed from the tail probabilities of the generated distributions.}}
\revision3{The stimuli are normalized Gaussian mixtures generated with varying component locations, spreads, and weights. Both encodings reuse the same fixed stimulus sequence and answers. T1 and T2 each contain five trials per subgoal. The tutorial explains density area and CCDF height with worked examples and revealed answers. Before each task category, participants familiarized themselves with example cases.}

\revision{\textbf{Experimental Setup and Statistical Analysis.}
A total of 12 participants were recruited for the study. Each participant completed all five tasks using both the CCDF and Density encodings. \revision3{The five task categories were presented in a fixed order. Within each category, participants completed the trials for one encoding as a block, followed by the trials for the other encoding, before proceeding to the next category. Encoding order was counterbalanced across participants: six used CCDF first and six used Density first.} \revision3{Ten trials were scheduled per task and encoding (100 per participant).
Participant means were computed from available trials without imputation.}
We recorded task completion time and accuracy for each task. For statistical analysis, we first averaged results within each participant for each task, and then conducted two-tailed paired \(t\)-tests on participant-level means. \revision3{Reported $p$ values use the Student $t$ distribution with 11 degrees of freedom and are unadjusted for multiple comparisons.} For the probability estimation task (T3), we computed accuracy as \(100-\text{absolute percentage-point error}\).}

\revision{\textbf{Results.}
As shown in \autoref{fig:perceptual_study}, CCDF achieved significantly higher accuracy in T1 (\textit{Extremes}, \(p < .05\)), T3 (\textit{Probability Estimation}, \(p < .001\)), T4 (\textit{Cluster Weight}, \(p < .001\)), and T5 (\textit{Entity Comparison}, \(p < .001\)); no significant difference was observed for T2 (\textit{Consistency}). For completion time, CCDF was significantly faster in T1 (\(p < .001\)) and T3 (\revision3{\(p = .002\)}), with no significant time differences in T2, T4, or T5.}

\subsection{User Study}
\label{sec:user_study}

\begin{figure*}[ht]
    \centering
    \includegraphics[width=\linewidth]{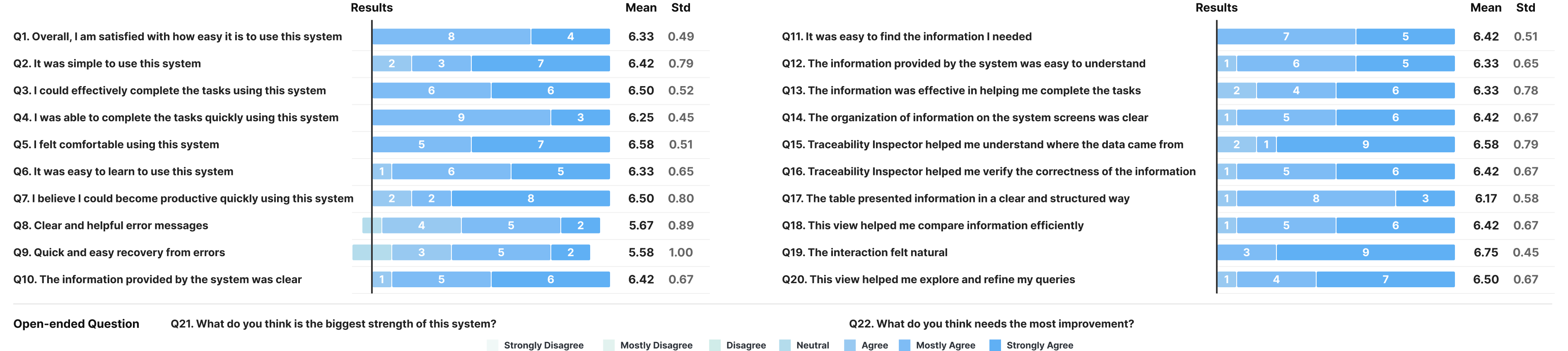}
    \caption{Quantitative results of the post-study questionnaire. The figure illustrates the score distributions, means, and standard deviations for the 20 closed-ended questions (Q1-Q20) evaluated on a 7-point Likert scale (1 = Strongly Disagree, 7 = Strongly Agree).
    }
    \label{fig:user_study}
\end{figure*}

To collect usability feedback from users with diverse but relevant backgrounds, we recruited 12 participants (P1--P12, comprising 8 males and 4 females) from a range of academic and professional fields, including data analysis, computer science, international trade, business engineering, and software development.
\revision{We treated participants with either relevant professional roles or at least three years of experience in visualization, data analysis, software development, or web-based information organization as expert users.
By this criterion, 7 participants (P1, P2, P3, P6, P7, P11, P12) were expert users.}
This cohort included two visualization experts (P3, P11), three data analysts (P1, P2, P7), two software developers (P6, P12), four graduate students (P4, P8, P9, P10), and one undergraduate student (P5).
All participants reported substantial experience in retrieving unstructured data from the Internet and performing data cleaning and analysis in their daily academic research or professional work.
We also collected participants' experience with data visualization tools; on average, they reported $4.08$ years of experience.
The study was conducted in a local experimental setting.
Participants used a 27-inch monitor and interacted with the system through a Chrome browser.

\textbf{Experimental Procedure.}
Each session lasted approximately 45 minutes.
Researchers first introduced the study background and the system workflow: users submit natural-language queries, and the system retrieves web evidence, structures numerical records, and aligns them for inspection.
Participants then completed a 10-minute tutorial covering the core interactions, including query input, backend crawling, cell-level inspection, uncertainty visualization, and provenance tracking.
After 10 minutes of free exploration, participants formulated multi-dimensional comparison queries based on their own interests or professional backgrounds.
Their queries covered diverse real-world scenarios, including cross-country housing prices, national land areas, macroeconomic price levels, new energy vehicle sales, and monthly active users (MAU) of popular games.


After submitting a query in the chat interface, the system first parsed the intent and prompted users to confirm the inferred retrieval scope.
\revision3{For queries with approximately ten entities and ten attributes (about 100 cells), observed wall-clock time was 1--2 minutes, from scope confirmation to completion of crawling, extraction, and aggregation for the entire table.
Cells run concurrently, and results are returned and displayed as soon as each cell finishes processing. Users can inspect completed cells while the remaining cells are still being processed. Additional cells increase processing and token consumption.}
\revision{The token cost of \toolname{} mainly comes from input evidence snippets rather than full web pages or long-form generation. For a query with \(m\) attributes and \(n\) entities, if each cell retains at most \(s\) sources and each evidence snippet contains \(k\) tokens on average, \revision3{an illustrative snippet-input estimate is} \(T_{in} \approx mnsk\). The output is usually short structured JSON; if each cell produces \(q\) output tokens on average, then \(T_{out} \approx mnq\). \revision3{Using illustrative rates of USD 2.50 per million input tokens and USD 15.00 per million output tokens, the corresponding token-cost component is} \(C \approx 2.5T_{in}/10^6 + 15T_{out}/10^6\). In practice, the same report may support multiple entities or attributes, and DataRecycle deduplicates retained sources across cells. If \(u\) denotes the number of unique retained evidence snippets after deduplication, the input scale is closer to \(uk\), where \(u \leq mns\). \revision3{This snippet-token estimate excludes search-service fees, crawling, and additional parsing, auditing, retry, and re-extraction calls.}}

After the data was retrieved, participants completed the following three steps: (1) \emph{Overview inspection:} participants examined the table view to understand the overall distribution, including both the CCDF bar encoding and the probability distribution visualization, and assessed whether the results could answer their query; (2) \emph{Focused exploration:} participants interacted with the table using the semantic lens to inspect specific cells in the table view; and (3) \emph{Verification and provenance tracking:} participants inspected individual cells using the traceability inspector and followed source links to examine data provenance.
After completing the tasks, participants completed a post-study questionnaire (see \autoref{fig:user_study}).
The questionnaire included items adapted from PSSUQ~\cite{lewis1992psychometric}.
It contained 20 closed-ended Likert-scale questions (Q1-Q20) evaluating system effectiveness, efficiency, and usability, as well as two open-ended questions (Q21, Q22) regarding \emph{useful features} (Q21) and \emph{suggestions for improvement} (Q22), respectively.

\textbf{Results.}
\revision3{Most questionnaire items had mean scores above 6.0 on the 7-point scale (\autoref{fig:user_study}). The item on natural interaction in the chat view had the highest mean (Q19, \(M=6.75\)). P12 noted that the unified table made it faster to compare sources and check values.}
\revision3{Ten participants identified provenance as useful in Q21. Participants rated the inspector as helpful for understanding data sources (Q15, \(M=6.58\)) and checking accuracy (Q16, \(M=6.42\)). P8 contrasted checking evidence directly in the inspector with manually extracting and verifying facts from long AI-generated answers.}
\revision3{We submitted 12 comparative factual queries to ChatGPT with web search enabled and checked the answers for factual accuracy, support from sources, and source coverage.
Six answers contained factual errors or claims that their sources did not support. Three others relied heavily on a single source or reproduced values from an existing table.}

\section{Discussion and Future Work}\label{sec:discussion}

\revision3{\textbf{Limitations.}
\toolname{} organizes numerical web evidence into comparable tables with links to the sources. Several limitations remain.
\textbf{1)~Evaluation scope.} The module benchmark measures extraction and auditing, the perceptual study tests how participants interpret the encodings, and the user study captures perceived usefulness. Even when retrieved sources agree, search rankings and limited retrieval depth can leave out relevant or contradictory reports. We did not compare evidence coverage, contradiction detection, or cost across retrieval depths. For source weighting, we tested how rankings change as the weights vary. We have yet to compare the scores with independent assessments of source reliability or compare verification outcomes under uniform or alternative weights. In the user study, we did not systematically assess whether participants correctly detected or corrected errors. We also did not compare final table accuracy, evidence coverage, or verification time in a controlled study.
\textbf{2)~Schema omissions.} Omitted attributes may leave no empty cells. Users can add them by refining the schema. The module benchmark uses manually specified targets and supplied texts, so it cannot assess retrieval coverage, initial schema completeness, or whether users find and recover omissions.
\textbf{3)~Source dependence.} Paragraph deduplication cannot identify dependent sources. Weighting republications and paraphrases separately can overstate support for the same evidence. Grouping dependent sources before weighting could reduce this effect.}

\textbf{Context-Aware Data via Local RAG.}
\toolname{} currently focuses on open-web data, while many organizations and individuals maintain large volumes of private ``dark data''~\cite{darkdata}, such as business logs, analytical reports, emails, and historical dashboards.
A promising direction is to incorporate local RAG mechanisms for structuring such private, heterogeneous repositories without exposing sensitive content to external LLM services.
This extension requires clear processing boundaries: sensitive data could be retrieved and modeled locally, whereas abstracted or sanitized representations could be passed to cloud-based LLMs for further reasoning.
Such local--cloud collaboration may help balance organizational context, analytical capability, and data security.

\textbf{Beyond Numerical Representations.}
\toolname{} currently focuses on numerical data, where facts can be aligned in a shared ontology space and presented in tabular form.
However, many real-world scenarios are better expressed through links and relationships among entities, forming relational structures that require conceptual disambiguation.
A future direction is to extend the ontology space beyond attribute-value pairs to represent entity relationships and graph-structured data.
This extension would support richer relational semantics, but also introduces challenges in modeling higher-order relationships and maintaining semantic alignment across heterogeneous sources.

\revision3{\textbf{Inspecting Potential Source Manipulation.}}
Recent research highlights \revision3{the risk of manipulating content retrieved by generative engines}~\cite{aggarwal2024geo,mochizuki2025exposing}.
In such attacks, malicious actors flood the web with manipulated content to intentionally contaminate the retrieval pool.
\revision3{Retrieved content can then lead to misleading answers}, for example, a scam phone number~\cite{deng2025llmsearchpoisoning}.
\revision3{The provenance and duplication views let users trace apparent agreement back to its sources and inspect repeated or conflicting reports. Future work could test whether displaying source relationships and publication timing helps analysts spot coordinated manipulation, such as sudden spikes in publication volume or clusters of highly repetitive reports.}

\section{Conclusion}\label{sec:conclusion}

This work addresses the challenge of transforming scattered web evidence into verifiable structured data for analysis.
\toolname{} adopts a schema-guided, top-down retrieval strategy that targets explicit table cells and makes coverage gaps visible, together with a provenance-centered design that makes the data lineage from raw snippets to final cells inspectable.
The evaluations indicate that \toolname{} supports evidence extraction and helps users inspect and interpret conflicts across sources.
By combining algorithmic trust estimation with interactive provenance inspection and schema refinement, \toolname{} helps users stay in control of the analytical process.
Looking ahead, several directions may extend the system.
Incorporating local retrieval-augmented generation could support context-aware structuring over private organizational data while respecting privacy constraints.
Extending the system beyond numerical tables could support richer relational representations in a broader ontology space.

\section*{Acknowledgment}
This project is supported by the Ministry of Education, Singapore, under its
Academic Research Fund Tier 1 (NTU Tier 1: RG104/25) and Academic Research Fund
Tier 2 (Proposal ID: T2EP20222-0049). Any opinions, findings and conclusions,
or recommendations expressed in this material are those of the author(s) and
do not reflect the views of the Ministry of Education, Singapore.

\bibliographystyle{IEEEtran}
\bibliography{references}

@string{TVCG = "IEEE Transactions on Visualization and Computer Graphics"}

@string{CHI = "Proceedings of the ACM Conference on Human Factors in Computing Systems"}

@string{WWW = "Proceedings of the ACM Web Conference"}

@string{UIST = "Proceedings of the ACM Symposium on User Interface Software and Technology"}

@string{KDD = "Proceedings of the ACM SIGKDD International Conference on Knowledge Discovery and Data Mining"}

@article{mochizuki2025exposing,
  title={Exposing Citation Vulnerabilities in Generative Engines},
  author={Mochizuki, Riku and Komatsu, Shusuke and Noguchi, Souta and Ataka, Kazuto},
  journal={arXiv preprint arXiv:2510.06823},
  year={2025}
}

@article{web_data_extraction,
author = {Ferrara, Emilio and De Meo, Pasquale and Fiumara, Giacomo and Baumgartner, Robert},
title = {Web data extraction, applications and techniques},
year = {2014},
issue_date = {November 2014},
publisher = {Elsevier Science Publishers B. V.},
address = {NLD},
volume = {70},
number = {C},
issn = {0950--7051},
url = {https://doi.org/10.1016/j.knosys.2014.07.007},
doi = {10.1016/j.knosys.2014.07.007},
journal = {Know.-Based Syst.},
month = nov,
pages = {301--323},
numpages = {23}
}

@article{machine_knowledge,
author = {Weikum, Gerhard and Dong, Xin Luna and Razniewski, Simon and Suchanek, Fabian},
title = {Machine Knowledge: Creation and Curation of Comprehensive Knowledge Bases},
year = {2021},
issue_date = {Jul 2021},
publisher = {Now Publishers Inc.},
address = {Hanover, MA, USA},
volume = {10},
number = {2--4},
issn = {1931-7883},
url = {https://doi.org/10.1561/1900000064},
doi = {10.1561/1900000064},
journal = {Found. Trends Databases},
month = jul,
pages = {108--490},
numpages = {390}
}

@inproceedings{aggarwal2024geo,
  title={{GEO}: Generative engine optimization},
  author={Aggarwal, Pranjal and Murahari, Vishvak and Rajpurohit, Tanmay and Kalyan, Ashwin and Narasimhan, Karthik and Deshpande, Ameet},
  doi = {10.1145/3637528.3671900},
  booktitle=KDD,
  pages={5--16},
  year={2024}
}

@misc{deng2025llmsearchpoisoning,
  title        = {When {AI} Recommends Scammers: New Attack Abuses {LLM} Indexing to Deliver Fake Support Numbers},
  author       = {Deng, Qi and Aurascape Aura Labs},
  year         = {2025},
  month        = dec,
  howpublished = {\url{https://aurascape.ai/llm-search-poisoning-fake-support-numbers/}},
  note         = {Accessed: 2026-03-29}
}

@ARTICLE{darkdata,
  author={Corallo, Angelo and Crespino, Anna Maria and Vecchio, Vito Del and Lazoi, Mariangela and Marra, Manuela},
  journal={IEEE Transactions on Engineering Management}, 
  title={Understanding and Defining Dark Data for the Manufacturing Industry}, 
  year={2023},
  volume={70},
  number={2},
  pages={700-712},
  doi={10.1109/TEM.2021.3051981}}

@inproceedings{lewis1992psychometric,
  title={Psychometric evaluation of the post-study system usability questionnaire: The {PSSUQ}},
  author={Lewis, James R},
  booktitle={Proceedings of the human factors society annual meeting},
  volume={36},
  number={16},
  pages={1259--1260},
  year={1992}
}

@book{iso1993guide,
  title={Guide to the Expression of Uncertainty in Measurement},
  author={ISO., I and OIML, BIPM},
  year={1993},
  publisher={Aenor Madrid, Spain}
}

@article{karunathilaka2026compendia,
  title={Compendia: Automated Visual Storytelling Generation from Online Article Collection},
  author={Karunathilaka, Manusha and Lei, Litian and Gao, Yiming and Wang, Yong and Li, Jiannan},
  journal=TVCG,
  year={2026},
  publisher={IEEE}
}

@ARTICLE{lu2020winglets,
  author={Lu, Min and Wang, Shuaiqi and Lanir, Joel and Fish, Noa and Yue, Yang and Cohen-Or, Daniel and Huang, Hui},
  journal=TVCG, 
  title={Winglets: Visualizing Association with Uncertainty in Multi-class Scatterplots}, 
  year={2020},
  volume={26},
  number={1},
  pages={770-779},
  doi={10.1109/TVCG.2019.2934811}}

@article{wang1996beyond,
author = {Wang, Richard Y. and Strong, Diane M.},
title = {Beyond accuracy: what data quality means to data consumers},
year = {1996},
issue_date = {March 1996},
publisher = {M. E. Sharpe, Inc.},
address = {USA},
volume = {12},
number = {4},
issn = {0742-1222},
doi = {10.1080/07421222.1996.11518099},
journal = {J. Manage. Inf. Syst.},
month = mar,
pages = {5--33},
numpages = {29}
}

@inproceedings{castillo2011information,
  title={Information credibility on twitter},
  author={Castillo, Carlos and Mendoza, Marcelo and Poblete, Barbara},
  booktitle={Proceedings of the 20th international conference on World wide web},
  pages={675--684},
  year={2011},
  doi={10.1145/1963405.1963500}
}

@article{dong2015knowledge,
  author       = {Xin Luna Dong and
                  Evgeniy Gabrilovich and
                  Kevin Murphy and
                  Van Dang and
                  Wilko Horn and
                  Camillo Lugaresi and
                  Shaohua Sun and
                  Wei Zhang},
  title        = {Knowledge-Based Trust: Estimating the Trustworthiness of Web Sources},
  journal      = {Proc. {VLDB} Endow.},
  volume       = {8},
  number       = {9},
  pages        = {938--949},
  year         = {2015},
  url          = {http://www.vldb.org/pvldb/vol8/p938-dong.pdf},
  doi          = {10.14778/2777598.2777603},
  bibsource    = {dblp computer science bibliography, https://dblp.org}
}

@article{li2016survey,
author = {Li, Yaliang and Gao, Jing and Meng, Chuishi and Li, Qi and Su, Lu and Zhao, Bo and Fan, Wei and Han, Jiawei},
title = {A Survey on Truth Discovery},
year = {2016},
issue_date = {December 2015},
publisher = {Association for Computing Machinery},
address = {New York, NY, USA},
volume = {17},
number = {2},
issn = {1931-0145},
url = {https://doi.org/10.1145/2897350.2897352},
doi = {10.1145/2897350.2897352},
journal = {SIGKDD Explor. Newsl.},
month = feb,
pages = {1--16},
numpages = {16}
}

@article{zhang2017deepdive,
author = {Zhang, Ce and R\'{e}, Christopher and Cafarella, Michael and De Sa, Christopher and Ratner, Alex and Shin, Jaeho and Wang, Feiran and Wu, Sen},
title = {DeepDive: declarative knowledge base construction},
year = {2017},
issue_date = {May 2017},
publisher = {Association for Computing Machinery},
address = {New York, NY, USA},
volume = {60},
number = {5},
issn = {0001-0782},
url = {https://doi.org/10.1145/3060586},
doi = {10.1145/3060586},
journal = {Commun. ACM},
month = apr,
pages = {93--102},
numpages = {10}
}

@article{pang1997approaches,
  title={Approaches to uncertainty visualization},
  author={Pang, Alex T and Wittenbrink, Craig M and Lodha, Suresh K},
  journal={The Visual Computer},
  volume={13},
  number={8},
  pages={370--390},
  year={1997},
  publisher={Springer},
  doi={10.1007/s003710050111}
}

@article{banko2007open,
    author = {Etzioni, Oren and Banko, Michele and Soderland, Stephen and Weld, Daniel S.},
    title = {Open information extraction from the web},
    year = {2008},
    issue_date = {December 2008},
    publisher = {Association for Computing Machinery},
    address = {New York, NY, USA},
    volume = {51},
    number = {12},
    url = {https://doi.org/10.1145/1409360.1409378},
    doi = {10.1145/1409360.1409378},
    journal = {Communications of the ACM},
    month = dec,
    pages = {68--74},
    numpages = {7}
}

@inproceedings{dong2014knowledge,
  title={Knowledge vault: a web-scale approach to probabilistic knowledge fusion},
  author={Dong, Xin and Gabrilovich, Evgeniy and Heitz, Geremy and Horn, Wilko and Lao, Ni and Murphy, Kevin and Strohmann, Thomas and Sun, Shaohua and Zhang, Wei},
  booktitle=KDD,
  pages={601--610},
  year={2014},
  doi={10.1145/2623330.2623623}
}

@inproceedings{agichtein2000snowball,
  title={Snowball: Extracting relations from large plain-text collections},
  author={Agichtein, Eugene and Gravano, Luis},
  booktitle={Proceedings of the fifth ACM conference on Digital libraries},
  pages={85--94},
  year={2000},
  doi={10.1145/336597.336644}
}

@inproceedings{ma2015faitcrowd,
  title={FaitCrowd: Fine Grained Truth Discovery for Crowdsourced Data Aggregation},
  author={Ma, Fenglong and Li, Yaliang and Li, Qi and Qiu, Minghui and Gao, Jing and Zhi, Shi and Su, Lu and Zhao, Bo and Ji, Heng and Han, Jiawei},
  booktitle=KDD,
  pages={745--754},
  year={2015},
  doi={10.1145/2783258.2783314}
}

@inproceedings{li2020towards,
  title={Towards Fair Truth Discovery from Biased Crowdsourced Answers},
  author={Li, Yanying and Sun, Haipei and Wang, Wendy Hui},
  booktitle=KDD,
  pages={399--409},
  year={2020},
  doi={10.1145/3394486.3403102}
}

@article{ye2022constrained,
  title={Constrained Truth Discovery},
  author={Ye, Chen and Wang, Hongzhi and Zheng, Kangjie and Kong, Youkang and Zhu, Rong and Gao, Jing and Li, Jianzhong},
  journal={IEEE Transactions on Knowledge and Data Engineering},
  volume={34},
  number={10},
  pages={4859--4872},
  year={2022},
  publisher={IEEE},
  doi={10.1109/TKDE.2020.2982393}
}

@article{song2019where,
  title={Where's my data? Evaluating visualizations with missing data},
  author={Song, Hayeong and Szafir, Danielle Albers},
  journal=TVCG,
  volume={25},
  number={1},
  pages={914--924},
  year={2019},
  publisher={IEEE},
  doi={10.1109/TVCG.2018.2864914}
}

@article{correll2014error,
  title={Error bars considered harmful: Exploring alternate encodings for mean and error},
  author={Correll, Michael and Gleicher, Michael},
  journal=TVCG,
  volume={20},
  number={12},
  pages={2142--2151},
  year={2014},
  publisher={IEEE},
  doi={10.1109/TVCG.2014.2346298}
}

@article{gershon1998visualization,
  title={Visualization of an imperfect world},
  author={Gershon, Nahum D},
  journal={IEEE Computer Graphics and Applications},
  volume={18},
  number={4},
  pages={43--45},
  year={1998},
  publisher={IEEE},
  doi={10.1109/38.689662}
}

@article{sanyal2009user,
  title={A user study to compare four uncertainty visualization methods for 1D and 2D datasets},
  author={Sanyal, Jibonananda and Zhang, Song and Bhattacharya, Gargi and Amburn, Phil and Moorhead, Robert J},
  journal=TVCG,
  volume={15},
  number={6},
  pages={1209--1218},
  year={2009},
  publisher={IEEE},
  doi={10.1109/TVCG.2009.114}
}

@article{athawale2024uncertainty,
  title={Uncertainty Visualization of Critical Points of 2D Scalar Fields for Parametric and Nonparametric Probabilistic Models},
  author={Athawale, Tushar M and Wang, Zhe and Gong, Qian and Klasky, Scott and Moreland, Kenneth and Pugmire, David and Johnson, Chris R and Rosen, Paul},
  journal=TVCG,
  doi={10.1109/TVCG.2024.3456393},
  year={2024},
  publisher={IEEE}
}

@inproceedings{huang2025structvizor,
  title={StructVizor: Interactive Profiling of Semi-Structured Textual Data},
  author={Huang, Y. and Miao, Y. and Weng, D. and Perer, A. and Wu, Y.},
  booktitle=CHI,
  pages={1--18},
  year={2025},
  publisher={ACM},
  doi={10.1145/3706598.3713484}
}

@article{lee2025hints,
  title={HINTs: Sensemaking on Large Collections of Documents With Hypergraph Visualization and INTelligent Agents},
  author={Lee, S. Y.-T. and Ma, K.-L.},
  journal=TVCG,
  volume={31},
  number={9},
  pages={5532--5546},
  year={2025},
  publisher={IEEE},
  doi={10.1109/TVCG.2024.3459961}
}

@inproceedings{tsai2025harmful,
  title={Harmful Terms and Where to Find Them: Measuring and Modeling Unfavorable Financial Terms and Conditions in Shopping Websites at Scale},
  author={Tsai, E. and Mangaokar, N. and Zheng, B. and Zheng, H. and Prakash, A.},
  booktitle=WWW,
  pages={990--1003},
  year={2025},
  publisher={ACM},
  doi={10.1145/3696410.3714573}
}

@article{ferreira2025vablueprint,
  title={VA-Blueprint: Uncovering Building Blocks for Visual Analytics System Design},
  author={Ferreira, L. and Moreira, G. and Miranda, F.},
  journal=TVCG,
  year={2025},
  publisher={IEEE},
  note={Early Access},
  doi={10.1109/TVCG.2025.3634809}
}

@article{chen2025conch,
  title={Conch: Competitive Debate Analysis via Visualizing Clash Points and Hierarchical Strategies},
  author={Chen, Q. and Wang, Y. and Yu, Y. and Zhu, X. and Yu, X. and Wang, R.},
  journal=TVCG,
  year={2025},
  publisher={IEEE},
  note={Early Access},
  doi={10.1109/TVCG.2025.3634629}
}

@inproceedings{li2025bridging,
  title={Bridging the Gap: Aligning Language Model Generation with Structured Information Extraction via Controllable State Transition},
  author={Li, H. and Ren, Y. and Cao, Y. and Li, Y. and Fang, F. and Lin, Z. and Wang, S.},
  booktitle= WWW,
  pages={1811--1821},
  year={2025},
  publisher={ACM},
  doi={10.1145/3696410.3714571}
}

@inproceedings{ni2025fgcibgc,
  title={FG-CIBGC: A Unified Framework for Fine-Grained and Class-Incremental Behavior Graph Classification},
  author={Ni, Z. and Fan, P. and Dai, S. and Zhang, B. and Wan, H. and Zhao, X.},
  booktitle=WWW,
  pages={1166--1181},
  year={2025},
  publisher={ACM},
  doi={10.1145/3696410.3714960}
}

@inproceedings{chen2025semantics,
  title={Semantics-Aware Cookie Purpose Compliance},
  author={Chen, B. and Lyu, J. and Wu, T. and Chhetri, M. B. and Bai, G.},
  booktitle=WWW,
  pages={1602--1613},
  year={2025},
  publisher={ACM},
  doi={10.1145/3696410.3714746}
}

@article{zhao2014fluxflow,
  title={{\#FluxFlow}: Visual Analysis of Anomalous Information Spreading on Social Media},
  author={Zhao, Jian and Cao, Nan and Wen, Zhen and Song, Yale and Lin, Yu-Ru and Collins, Christopher},
  journal=TVCG,
  volume={20},
  number={12},
  pages={1773--1782},
  year={2014},
  publisher={IEEE}
}

@article{cao2015targetvue,
  title={Targetvue: Visual analysis of anomalous user behaviors in online communication systems},
  author={Cao, Nan and Shi, Conglei and Lin, Sabrina and Lu, Jie and Lin, Yu-Ru and Lin, Ching-Yung},
  journal=TVCG,
  volume={22},
  number={1},
  pages={280--289},
  year={2015},
  publisher={IEEE}
}

@inproceedings{bors2014qualitytrails,
  title={Qualitytrails: Data quality provenance as a basis for sensemaking},
  author={Bors, Christian and Gschwandtner, Theresia and Miksch, Silvia and G{\"a}rtner, Johannes},
  booktitle={Proceedings of the International Workshop on Analytic Provenance for Sensemaking},
  year={2014}
}

@ARTICLE{Ragan2016CPVDA,
  author={Ragan, Eric D. and Endert, Alex and Sanyal, Jibonananda and Chen, Jian},
  journal=TVCG, 
  title={Characterizing Provenance in Visualization and Data Analysis: An Organizational Framework of Provenance Types and Purposes}, 
  year={2016},
  volume={22},
  number={1},
  pages={31-40},
  doi={10.1109/TVCG.2015.2467551}}

@ARTICLE{7536106,
  author={Dasgupta, Aritra and Lee, Joon-Yong and Wilson, Ryan and Lafrance, Robert A. and Cramer, Nick and Cook, Kristin and Payne, Samuel},
  journal=TVCG, 
  title={Familiarity Vs Trust: A Comparative Study of Domain Scientists' Trust in Visual Analytics and Conventional Analysis Methods}, 
  year={2017},
  volume={23},
  number={1},
  pages={271-280},
  doi={10.1109/TVCG.2016.2598544}}

@article{https://doi.org/10.1111/cgf.14034,
author = {Chatzimparmpas, A. and Martins, R. M. and Jusufi, I. and Kucher, K. and Rossi, F. and Kerren, A.},
title = {The State of the Art in Enhancing Trust in Machine Learning Models with the Use of Visualizations},
journal = {Computer Graphics Forum},
volume = {39},
number = {3},
pages = {713-756},
doi = {https://doi.org/10.1111/cgf.14034},
url = {https://onlinelibrary.wiley.com/doi/abs/10.1111/cgf.14034},
eprint = {https://onlinelibrary.wiley.com/doi/pdf/10.1111/cgf.14034},
year = {2020}
}

@ARTICLE{6327271,
  author={Cao, Nan and Lin, Yu-Ru and Sun, Xiaohua and Lazer, David and Liu, Shixia and Qu, Huamin},
  journal=TVCG, 
  title={Whisper: Tracing the Spatiotemporal Process of Information Diffusion in Real Time}, 
  year={2012},
  volume={18},
  number={12},
  pages={2649-2658},
  doi={10.1109/TVCG.2012.291}}

@article{MacEachren01012005,
author = {Alan M. MacEachren and Anthony Robinson and Susan Hopper and Steven Gardner and Robert Murray and Mark Gahegan and Elisabeth Hetzler},
title = {Visualizing Geospatial Information Uncertainty: What We Know and What We Need to Know},
journal = {Cartography and Geographic Information Science},
volume = {32},
number = {3},
pages = {139--160},
year = {2005},
publisher = {Taylor \& Francis},
doi = {10.1559/1523040054738936},
URL = {https://doi.org/10.1559/1523040054738936},
eprint = {https://doi.org/10.1559/1523040054738936}
}

@inproceedings{potter_quantification_2012,
	address = {Berlin, Heidelberg},
	title = {From {Quantification} to {Visualization}: {A} {Taxonomy} of {Uncertainty} {Visualization} {Approaches}},
	isbn = {978-3-642-32677-6},
	booktitle = {Uncertainty {Quantification} in {Scientific} {Computing}},
	publisher = {Springer Berlin Heidelberg},
	author = {Potter, Kristin and Rosen, Paul and Johnson, Chris R.},
	editor = {Dienstfrey, Andrew M. and Boisvert, Ronald F.},
	year = {2012},
	pages = {226--249},
}

@ARTICLE{6327283,
  author={Boukhelifa, Nadia and Bezerianos, Anastasia and Isenberg, Tobias and Fekete, Jean-Daniel},
  journal=TVCG, 
  title={Evaluating Sketchiness as a Visual Variable for the Depiction of Qualitative Uncertainty}, 
  year={2012},
  volume={18},
  number={12},
  pages={2769-2778},
  doi={10.1109/TVCG.2012.220}}

@ARTICLE{7536212,
  author={Correll, Michael and Heer, Jeffrey},
  journal=TVCG, 
  title={Surprise! Bayesian Weighting for De-Biasing Thematic Maps}, 
  year={2017},
  volume={23},
  number={1},
  pages={651-660},
  doi={10.1109/TVCG.2016.2598618}}

@inproceedings{10.1145/2858036.2858558,
    author = {Kay, Matthew and Kola, Tara and Hullman, Jessica R. and Munson, Sean A.},
    title = {When (ish) is My Bus? User-centered Visualizations of Uncertainty in Everyday, Mobile Predictive Systems},
    year = {2016},
    isbn = {9781450333627},
    address = {New York, NY, USA},
    url = {https://doi.org/10.1145/2858036.2858558},
    doi = {10.1145/2858036.2858558},
    booktitle = CHI,
    pages = {5092--5103},
    numpages = {12},
    location = {San Jose, California, USA},
}

@inproceedings{10.1145/3173574.3173718,
author = {Fernandes, Michael and Walls, Logan and Munson, Sean and Hullman, Jessica and Kay, Matthew},
title = {Uncertainty Displays Using Quantile Dotplots or CDFs Improve Transit Decision-Making},
year = {2018},
isbn = {9781450356206},
publisher = {Association for Computing Machinery},
address = {New York, NY, USA},
url = {https://doi.org/10.1145/3173574.3173718},
doi = {10.1145/3173574.3173718},
booktitle = CHI,
pages = {1--12},
numpages = {12},
location = {Montreal QC, Canada},
series = {CHI '18}
}

@ARTICLE{7192667,
  author={Gschwandtnei, Theresia and B{\"o}gl, Markus and Federico, Paolo and Miksch, Silvia},
  journal=TVCG, 
  title={Visual Encodings of Temporal Uncertainty: A Comparative User Study}, 
  year={2016},
  volume={22},
  number={1},
  pages={539-548},
  doi={10.1109/TVCG.2015.2467752}}

@inproceedings{10.1145/3173574.3174216,
author = {Correll, Michael and Moritz, Dominik and Heer, Jeffrey},
title = {Value-Suppressing Uncertainty Palettes},
year = {2018},
isbn = {9781450356206},
publisher = {Association for Computing Machinery},
address = {New York, NY, USA},
url = {https://doi.org/10.1145/3173574.3174216},
doi = {10.1145/3173574.3174216},
booktitle = CHI,
pages = {1--11},
numpages = {11},
location = {Montreal QC, Canada},
}

@ARTICLE{8807299,
  author={Spinner, Thilo and Schlegel, Udo and Sch{\"a}fer, Hanna and El-Assady, Mennatallah},
  journal=TVCG, 
  title={expl{AI}ner: A Visual Analytics Framework for Interactive and Explainable Machine Learning}, 
  year={2020},
  volume={26},
  number={1},
  pages={1064-1074},
  doi={10.1109/TVCG.2019.2934629}}

@ARTICLE{9555810,
  author={Cheng, Furui and Liu, Dongyu and Du, Fan and Lin, Yanna and Zytek, Alexandra and Li, Haomin and Qu, Huamin and Veeramachaneni, Kalyan},
  journal=TVCG, 
  title={{VBridge}: Connecting the Dots Between Features and Data to Explain Healthcare Models}, 
  year={2022},
  volume={28},
  number={1},
  pages={378-388},
  doi={10.1109/TVCG.2021.3114836}}

@inproceedings{10.1145/2807442.2807478,
author = {Gao, Tong and Dontcheva, Mira and Adar, Eytan and Liu, Zhicheng and Karahalios, Karrie G.},
title = {{DataTone}: Managing Ambiguity in Natural Language Interfaces for Data Visualization},
year = {2015},
isbn = {9781450337793},
publisher = {Association for Computing Machinery},
address = {New York, NY, USA},
url = {https://doi.org/10.1145/2807442.2807478},
doi = {10.1145/2807442.2807478},
booktitle = UIST,
pages = {489--500},
numpages = {12},
location = {Charlotte, NC, USA},

}

@inproceedings{10.1145/2984511.2984588,
author = {Setlur, Vidya and Battersby, Sarah E. and Tory, Melanie and Gossweiler, Rich and Chang, Angel X.},
title = {Eviza: A Natural Language Interface for Visual Analysis},
year = {2016},
isbn = {9781450341899},
publisher = {Association for Computing Machinery},
address = {New York, NY, USA},
url = {https://doi.org/10.1145/2984511.2984588},
doi = {10.1145/2984511.2984588},
booktitle = UIST,
pages = {365--377},
numpages = {13},
location = {Tokyo, Japan},
}

@INPROCEEDINGS{8031599,
  author={Ren, Donghao and Brehmer, Matthew and Bongshin Lee and H{\"o}llerer, Tobias and Choe, Eun Kyoung},
  booktitle={Proceedings of the IEEE Pacific Visualization Symposium}, 
  title={{ChartAccent}: Annotation for data-driven storytelling}, 
  year={2017},
  volume={},
  number={},
  pages={230-239},
  doi={10.1109/PACIFICVIS.2017.8031599}
  }

@INPROCEEDINGS{ortona2018robust,
  author={Ortona, Stefano and Meduri, Venkata Vamsikrishna and Papotti, Paolo},
  booktitle={Proceedings of the IEEE International Conference on Data Engineering}, 
  title={Robust Discovery of Positive and Negative Rules in Knowledge Bases}, 
  year={2018},
  pages={1168--1179},
  doi={10.1109/ICDE.2018.00108}}

@inproceedings{Kaghazgaran2018TwoFace,
author = {Kaghazgaran, Parisa and Caverlee, James and Squicciarini, Anna},
title = {Combating Crowdsourced Review Manipulators: A Neighborhood-Based Approach},
year = {2018},
isbn = {9781450355810},
publisher = {Association for Computing Machinery},
address = {New York, NY, USA},
url = {https://doi.org/10.1145/3159652.3159726},
doi = {10.1145/3159652.3159726},
booktitle = {Proceedings of the ACM International Conference on Web Search and Data Mining},
pages = {306--314},
numpages = {9},
location = {Marina Del Rey, CA, USA},
series = {WSDM '18}
}

@article{wang2025scidasynth,
  author = {Wang, Xingbo and Huey, Samantha L. and Sheng, Rui and Mehta, Saurabh and Wang, Fei},
  title = {{SciDaSynth}: Interactive Structured Data Extraction From Scientific Literature With Large Language Model},
  journal = {Campbell Systematic Reviews},
  year = {2025},
  volume = {21},
  number = {4},
  pages = {e70073},
  doi = {10.1002/cl2.70073}
}

@article{zhang2024chartifytext,
  author = {Zhang, Songheng and Wang, Lei and Li, Toby Jia-Jun and Shen, Qiaomu and Cao, Yixin and Wang, Yong},
  title = {{ChartifyText}: Automated Chart Generation from Data-Involved Texts via {LLM}},
  journal = {arXiv preprint arXiv:2410.14331},
  year = {2024},
  doi = {10.48550/arXiv.2410.14331}
}

@article{li2026fox,
  author = {Li, Yikai and Wang, Yong},
  title = {{FOX}: Visual Exploration of Data Fact Outliers},
  journal = {arXiv preprint arXiv:2608.08671},
  year = {2026},
  doi = {10.48550/arXiv.2608.08671}
}

\vspace{-33px}
\begin{IEEEbiography}[{\includegraphics[width=1in,height=1.25in,clip,keepaspectratio]{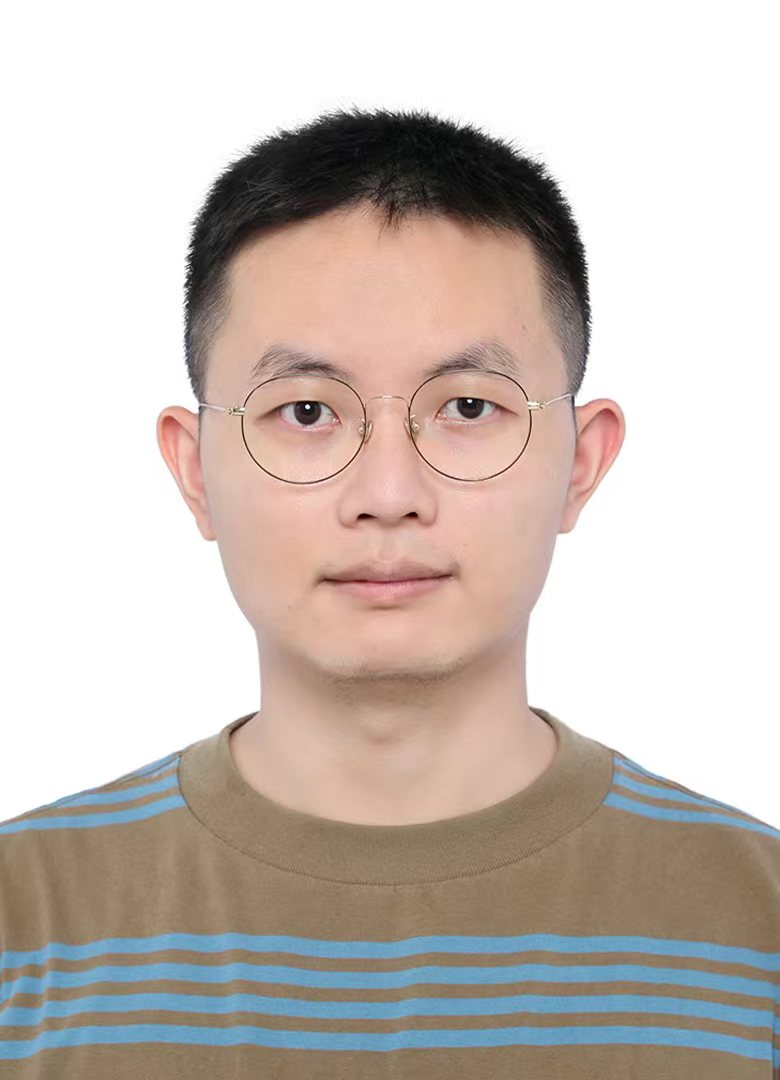}}]{Can Liu} is an assistant professor at the Central University of Finance and Economics.
He was previously a research fellow in the College of Computing and Data Science at Nanyang Technological University.
He received his Ph.D. in Intelligent Science (2023), B.Sc. in Computer Science (2018), and B.Ec in Economics (2018) degrees from Peking University.
His research interests lie in the field of deep learning-driven visualization, especially intelligent interaction for visualization. Refer to \url{https://vvoliucano.github.io} for more details.
\end{IEEEbiography}
\vspace{-33px}

\begin{IEEEbiography}[{\includegraphics[width=1in,height=1.20in,clip,keepaspectratio]{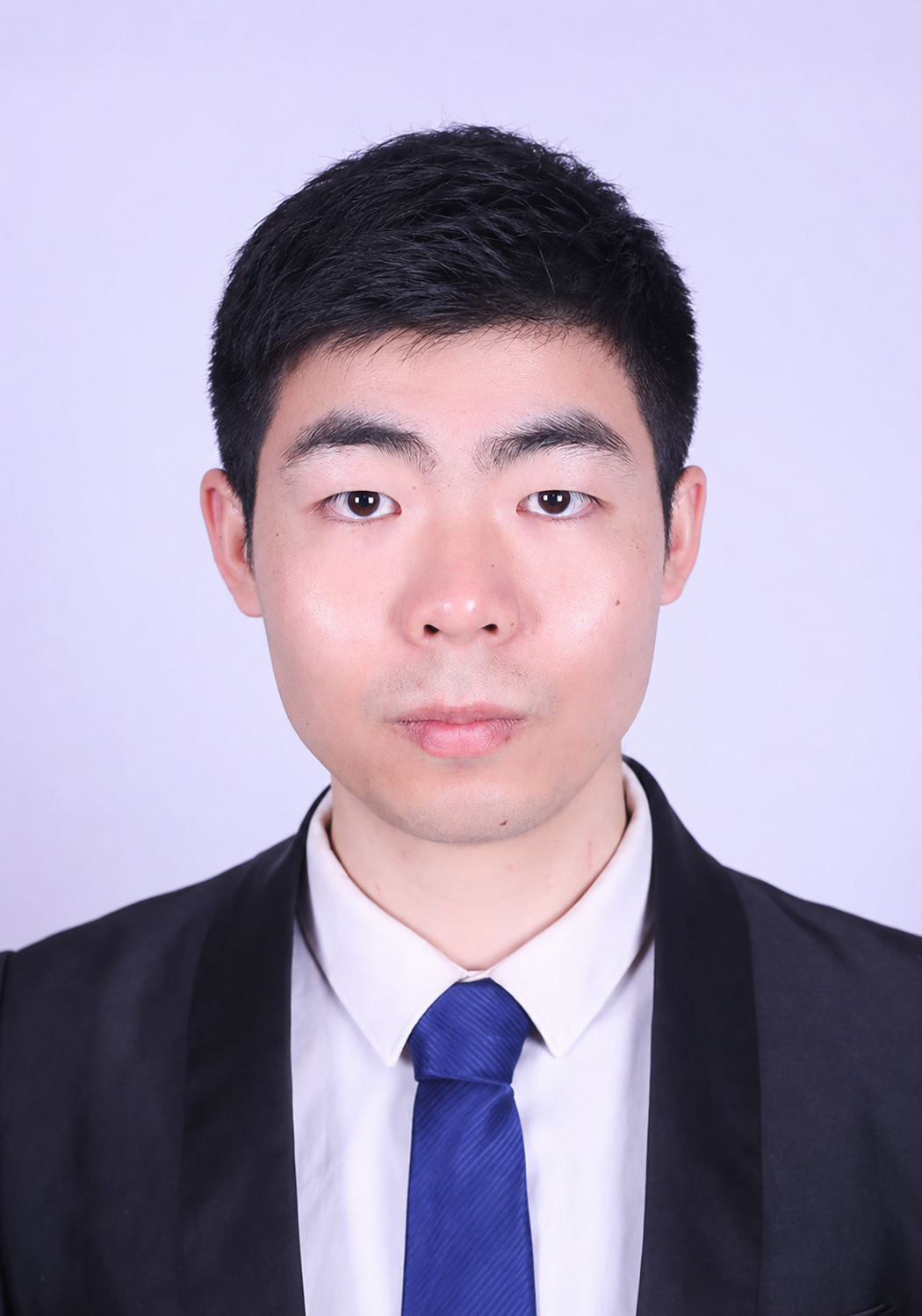}}]{Tengteng Zhu}
received the B.S. degree from North west Agriculture and Forestry University in 2017, and the M.S. degree from Beijing Information Science and Technology University in 2021. He is currently pursuing a Ph.D. degree at the Beijing Institute of Technology.
\end{IEEEbiography}
\vspace{-33px}

\begin{IEEEbiography}[{\includegraphics[width=1in,height=1.25in,clip,keepaspectratio]{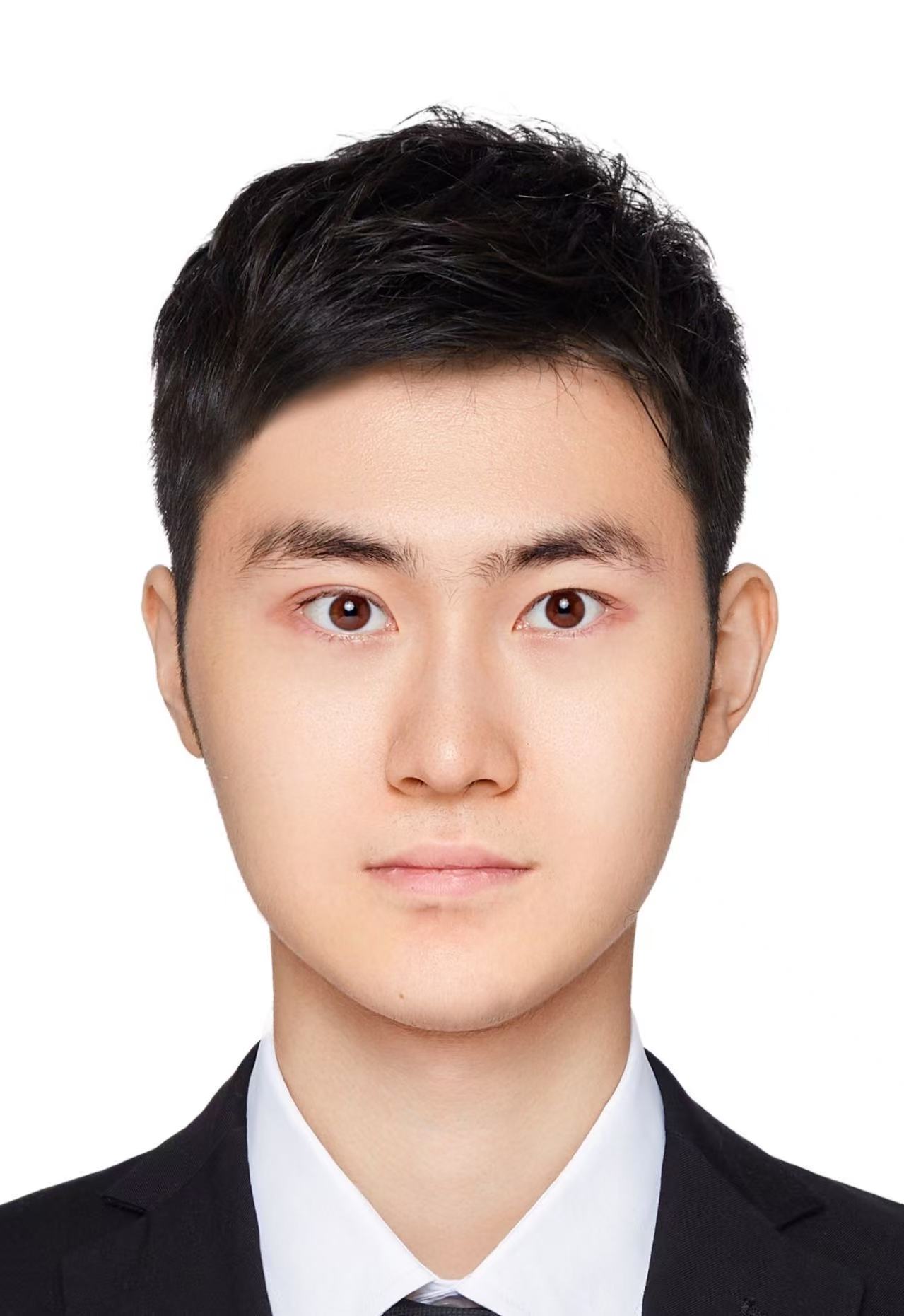}}]{Xinyu Jiang}
is a Master of Engineering (Software Engineering) at the University of Sydney, Australia. He received his B.Sc. degree in Software Engineering from Beijing Institute of Technology. His research interests include intelligent interactive systems, multimodal AI, and visual analytics.
\end{IEEEbiography}
\vspace{-33px}

\begin{IEEEbiography}[{\includegraphics[width=1in,height=1.25in,clip,keepaspectratio]{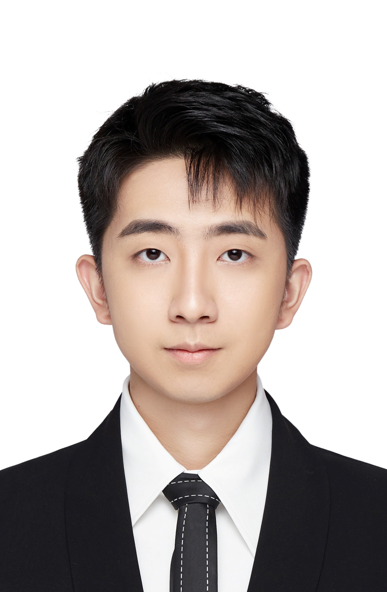}}]{Xiaolin Wen}
is currently a Ph.D. student in the College of Computing and Data Science, Nanyang Technological University (NTU). 
His research interests include on visualization for FinTech and LLM-assisted design study.
He received his master's degree in Computer Science and Technology from Sichuan University in 2023 and his dual bachelor's degree in Computer Science and Financial Engineering from Sichuan University in 2016.
For more information, kindly visit \url{https://wenxiaolin.com/}.
\end{IEEEbiography}

\vspace{-33px}


\begin{IEEEbiography}[{\includegraphics[width=1in,height=1.25in,clip,keepaspectratio]{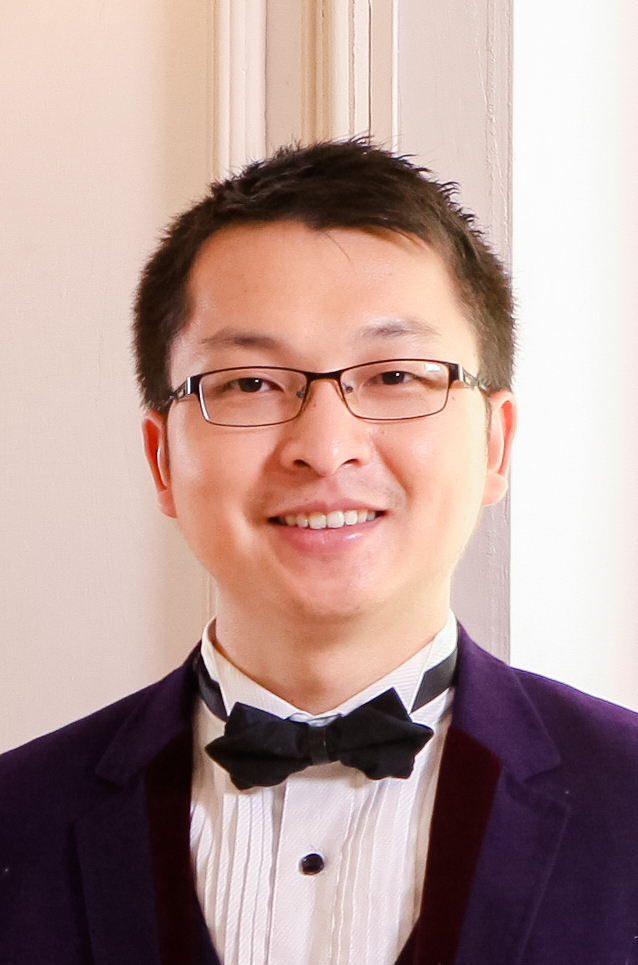}}]{Yong Wang}
is currently an assistant professor in the College of Computing and Data Science, Nanyang Technological University. Before that, he worked
as an assistant professor at Singapore Management University from 2020 to 2024. His research interests include data visualization, HCI, and human-AI collaboration, with an emphasis on their application to FinTech, quantum computing and online learning.
He obtained his Ph.D. in Computer Science from Hong Kong University of Science and Technology. He received his B.E. and M.E. from Harbin Institute of Technology and Huazhong University of Science and Technology, respectively. For more details, please refer to \url{http://yong-wang.org}.

\end{IEEEbiography}


\end{CJK}
\end{document}